\documentstyle[amssymb,latexsym,graphics,psfig,natbib209]{mn-nat}

\title[Flattening and truncation of stellar discs in edge-on spiral galaxies]{Flattening and
truncation of stellar discs in edge-on spiral galaxies}
\author[Michiel Kregel, Pieter C. van der Kruit, and Richard de Grijs]{Michiel
Kregel$^{1}$\thanks{E-mail: kregel@astro.rug.nl}, Pieter C. van der
Kruit$^{1}$, and Richard de Grijs$^{2,3}$
\\
$^1$ Kapteyn Astronomical Institute, University of Groningen, PO Box
800, 9700 AV Groningen, the Netherlands\\
$^2$ Astronomy Department, University of Virginia, PO Box 3818,
Charlottesville, VA 22903, USA\\
$^3$ Institute of Astronomy, University of Cambridge, Madingley
Road, Cambridge CB3 0HA\\
}

\date{Accepted 2002 April 4. Received 2001 June 1}
\pubyear{2001}

\begin{document}

\maketitle

\begin{abstract}
We analyse the global structure of the old stellar discs in 34 edge-on
spiral galaxies. The radial and vertical exponential scale parameters
of the discs are obtained by applying an improved two-dimensional
decomposition technique to our {\it I}-band photometry. We find a
clear increase in the disc scaleheight with maximum rotational
velocity, in accordance with observations of the stellar velocity
dispersions in galaxy discs. The range and maximum of the intrinsic
flattening of the disc light seem to increase with both maximum
rotational velocity and total H\,{\sc i} mass. We use the disc flattening
to estimate the disc contribution to the maximum rotational velocity,
resulting in an average of 57$\pm$22 percent. The disc light
distributions are further investigated for the presence of radial
truncations. We find that the radial light distributions of at least
20 spirals are truncated, corresponding to 60 percent of the
sample. For small scalelength spirals, which are the most numerous in
the local Universe, the results suggest that the average ratio of disc 
truncation radius to disc scalelength is at least four.
\end{abstract}


\section{Introduction}
\label{sec:intro}

The study of the light and mass distribution of galactic discs is of
fundamental importance for our understanding of the formation and
evolution of spiral galaxies. It is well established that globally,
the radial light distribution of galactic discs is exponential in
nature \citep{F70}, over a wide range of central surface brightness
(\citealt{JIII96}; \citealt*{BHB95}). This radial exponential
distribution can be understood in the framework of the collapse theory
of disc galaxy formation if the stellar mass-to-light ratio is
approximately constant with radius (\citealt{FE80,K87};
\citealt*{DSS97}). Perpendicular to the galactic plane, the light
distribution of the disc is also approximately exponential
(\citealt{KS81a,KS81b,KS82a}, hereafter KS1--3; \citealt*{WKH89};
\citealt{SG90}; \citealt*{GPK97}). While the radial structure
of the stellar disc is probably determined during disc galaxy
formation, the vertical structure largely reflects the secular disc
evolution \citep{W77,B84b,L84}. Together, the radial and vertical disc
structure determine the intrinsic disc flattening. Accordingly,
disc flattening is governed by both disc galaxy formation and
evolution and may offer additional insight into the physics of these.

The disc flattening can only be studied accurately in edge-on
galaxies, where the radial and vertical distributions are essentially
independent (KS1). Previous studies of large samples of edge-on
disc galaxies pointed out that, on average, the disc light
distribution slowly flattens from earlier to later morphological type
\citep{Gu92,K94}. These studies adopted a statistical approach, using
isophotal axial ratios. A similar trend is observed in a sample of 47
edge-ons \citep{G98}, using the ratio of the exponential scalelength
to the exponential scaleheight of the disc light \citep[see
also][]{SD00}. In this sample, a further comparison of the disc
flattening with other global parameters, such as rotational velocity
or H\,{\sc i} content, may reveal additional trends concerning disc
galaxy formation and evolution. Besides this, the flattening of the
disc light can be used to estimate the disc contribution to the
rotation curve \citep{B93}. This allows one to study the dark matter
content of spiral galaxies and provides an independent test of the
so-called maximum disc hypothesis \citep{A85,F92}. 

A related study of the stellar kinematics of galactic discs by
\citet{B93} shows that more massive spirals tend to have larger
stellar velocity dispersions. Together with the increase of the
stellar velocity dispersion with age \citep{W77}, this suggests that
more massive discs are more dynamically evolved. If true, then more
massive discs are also expected to have larger scaleheights. This can
easily be verified observationally in edge-on spiral galaxies.

Another global parameter describing the radial light distribution of
discs is its outer edge. This truncation of the disc light occurs at a
radius $R_{\rm max}$, which varies from 3 to 5 scalelengths among disc
galaxies \citep{K01}. It is most easily found in edge-on spirals
because of their higher surface brightness (KS1--3;
\citealt{BD94,PDLS00,SD00}; \citealt*{GKW01}). In less inclined
spirals, for which azimuthally averaged radial light profiles are
often studied, the non-axisymmetric component (e.g. due to spiral
structure) tends to smooth out a truncation present in the old
disc. This effect was first noted by \citet{K88}; for 16 face-on
spirals, of which 15 did not show any sign of a truncation in 
the radial light profile, he found that the three outermost isophotes
were much more closely spaced than the inner ones, providing clear
evidence for truncations. The origin of the truncation is still
unclear. If the truncation is also present in the disc mass
distribution, it may be caused by the angular momentum distribution of
the proto galaxy \citep{K87}. This would imply that H\,{\sc i}
observed beyond the stellar disc \citep[e.g.][]{BR97} has been
accreted \citep[cf.][]{B96}. Another possibility is that the
truncation corresponds to the radius at which the gas density drops
below a threshold density necessary for star formation
(\citealt{FE80}; KS3; \citealt*{SSE84}; \citealt{Ken89}).

In a previous study, \citet{G98} analysed {\it BVIJK$'$} photometry
of a statistically complete sample of nearby edge-on disc galaxies.
He used profiles taken at suitably chosen positions to determine both
the radial and vertical scale parameters (hereafter 1D method). The
determination of the radial scalelengths was based on radial profiles,
taken at some distance above the plane to minimize extinction. Still,
the uncertainty in this analysis is considerable. First, spiral
galaxies may have a surface brightness distribution which shows
significant local deviations from assumed smooth model distributions
\citep{SG90,J96}. Since the 1D method uses a small portion of the 
galaxy, it is susceptible to these local deviations. Secondly, the 1D
method does not consider the presence of a radial truncation. Finally,
an exponential was used as a fitting function, thereby neglecting the
effect of line of sight projection. These omissions have probably
caused a serious error in the disc scalelengths.

Here, we re-analyse the {\it I}-band photometry of a sub-sample of de
Grijs's sample, consisting of 34 spiral galaxies. In the first part of
this paper, we use a two-dimensional (2D) least-squares algorithm to
study the exponential scale parameters, and thereby the disc
flattening. This 2D method minimizes the effects of a radial
truncation and takes the line of sight projection into account. In
addition, it allows for a bulge-disc decomposition and adopts a mask
to minimize the effect of dust extinction. To assess the effect of
dust extinction outside this mask, the {\it B}-band data are subjected
to the same analysis. The 2D method was briefly introduced by
\citet{GKW01}, where the results of a detailed study of the radial
truncations of four spiral galaxies were presented. In the second part
of this paper we report on a similar study of the radial truncations
for our sample in the {\it I} band. Finally, the observed truncation
radii are compared to those predicted by the scenarios proposed for
the origin of the truncation. This paper is organized as follows. In
Sect.~\ref{sec:sample} we present the sample. The 2D least-squares
method is discussed and tested in Sect.~\ref{sec:sb_method}. In
Sect.~\ref{sec:results}, the results of its application to the {\it
I}-band photometry are presented, together with the results of our
study of the disc truncations. We further analyse and discuss the
results in Sect.~\ref{sec:discussion}. Our main findings are
summarized in Sect.~\ref{sec:conclusions}.

\section{The Sample}
\label{sec:sample}

The parent sample \citep{G98} contains 47 spirals and lenticulars
and was selected from the ESO-LV catalogue \citep{LV89} according to
the following criteria:

\begin{tabular}{ll}
(i)   & inclination, {\it i} $\geq$ 87$^{\circ}$; \\
(ii)  & blue diameter, {\it D}$^{\it B}_{25}$ $>$ 2$'$.2; \\
(iii) & Hubble type ranging from S0--Sd;\\
(iv)  & non-interacting\\
\end{tabular}

\noindent
From this sample we selected all regular spiral galaxies and excluded
those that are clearly lopsided or warped. Important properties of the
34 selected spirals are summarized in Table~\ref{tab:dfit2d_disc},
columns (1)--(6) \footnote{All distance dependent parameters in 
this paper are calculated using ${\it H}_{0}$=75 \mbox{km s$^{-1}$}
Mpc$^{-1}$}. We note that although the sub-sample covers a large
range in maximum rotational velocities ($v_{\rm max}$ = 50--400 km
s$^{-1}$), it is dominated by spirals of intermediate- to late-type.

We checked the completeness of the sub-sample with the 
${\it V}/{\it V}_{\rm max}$ test \citep{D90}. Based on the blue
angular diameters given by the ESO-LV catalogue, the average ${\it
V}/{\it V}_{\rm max}$ is  $0.58\ \pm\ 0.06$. This is slightly larger
than the average of 0.5 expected for a randomly distributed sample. In
the remainder of this paper, one should keep in mind that our sample
is of modest size and may not be statistically complete.

While our {\it J} and {\it K$'$}-band data more closely resemble the
underlying disc mass distribution than the optical {\it BVI} data
\citep{RR93}, the quality of the former is poorer \citep{G98} and will
not be used. Of the optical passbands, the light in the {\it I} band
is the least affected by dust extinction and the most dominated by the
old stellar population \citep[e.g.][]{CB91}. We will therefore
primarily use those data to study the global structure of stellar
discs. The data reduction and sky subtraction have been discussed by
\citet{G97,G98}.

\section{The two dimensional method}
\label{sec:sb_method}

\subsection{The model}
\label{sec:sb_model}

The spiral galaxy luminosity density is assumed to be composed of a
disc and a bulge component. The disc is assumed to be axisymmetric,
transparent and, in first instance, without a radial truncation:

\begin{equation}
L(R, z) = L_{0}\ e^{-R/h_{\rm R}} e^{-z/h_{\rm z}},
\label{eqn:lumdisc}
\end{equation}

\noindent
where ({\it R}, {\it z}) are the cylindrical coordinates, $L_{0}$ is
the central luminosity density and $h_{\rm R}$ and $h_{\rm z}$ are the
radial scalelength and the vertical scaleheight (in terms of the
vertical scale parameter $z_{0}$; $h_{\rm z}$ = 0.5 $z_{0}$). While 
other functionalities have been proposed for the vertical
distribution, most notably the locally isothermal (KS1) and sech$(z)$
distributions \citep{K88}, the exact choice is unimportant for our
purpose (Sect.~\ref{sec:sb_approach}). The vertical exponential
behaviour is taken to be independent of radius following
\citet{GP97}. They showed that, in the discs of most of the 
spirals in our sample (at least all late-type galaxies, which is about
two thirds), the scaleheight does not increase with radius. It is
further assumed that the galaxies are perfectly edge-on. As KS1
demonstrate, the error of the scale parameters due to 
this assumption is negligible as long as {\it i} $\geq$ 86$^{\circ}$
(see also \citealt{GPK97}), which is the case for all the
galaxies in our sample. With these assumptions the disc surface
brightness distribution is then given by (KS1):

\begin{equation}
\Sigma_{\rm disc}(R', z) = \Sigma_{0}\ (R'/h_{\rm R})\ K_{1}(R'/h_{\rm R})\ e^{-z/h_{\rm z}},
\label{eqn:disc}
\end{equation}

\noindent
where {\it R}$'$ is the projected radius along the major axis,
$\Sigma_{0} = 2 h_{\rm R} L_{0}$ is the projected {\it edge-on}
central surface brightness and $K_{1}$ is the modified Bessel function
of the first order.

The surface brightness distribution of the bulge is modelled either by
an exponential:

\begin{equation}
\Sigma_{\rm bulge}(R', z) = 5.360\ \Sigma_{\rm e}\ e^{-1.679\ r/r_{\rm e}},
\label{eqn:bulge_exp}
\end{equation}

\noindent
or by an {\it r}$^{1/4}$ law:

\begin{equation}
\Sigma_{\rm bulge}(R', z) = \Sigma_{\rm e}\ e^{-7.669\ [(r/r_{\rm e})^{1/4} - 1]},
\label{eqn:bulge_dev}
\end{equation}

\noindent
here $r$ is the projected radius; $r = \sqrt{R'^{2}+(z/q)^{2}}$,
$r_{\rm e}$ is the effective radius, $\Sigma_{\rm e}$ the effective
surface brightness, and {\it q} the bulge axial ratio (the ratio of
the bulge minor to major axis diameter). The best-fitting bulge, in
the reduced $\chi^{2}$ sense, is adopted. To account for the effect of
seeing, the model bulges are convolved with a two-dimensional Gaussian
PSF. For the disc component the seeing effect can be neglected
(Sect.~\ref{sec:sb_tests}). The central positions and position angles
of the galaxies, determined by fitting ellipses to the {\it I}-band
images (Sect.~{\ref{sec:sb_results}}), are fixed parameters in the
fit. Thus, the adopted model has at most six free parameters;
$\Sigma_{0}$, $h_{\rm R}$, $h_{\rm z}$, $\Sigma_{\rm e}$, $r_{\rm e}$,
and $q$.

\begin{figure*}
\centering
	\resizebox{17.5cm}{!}{\includegraphics{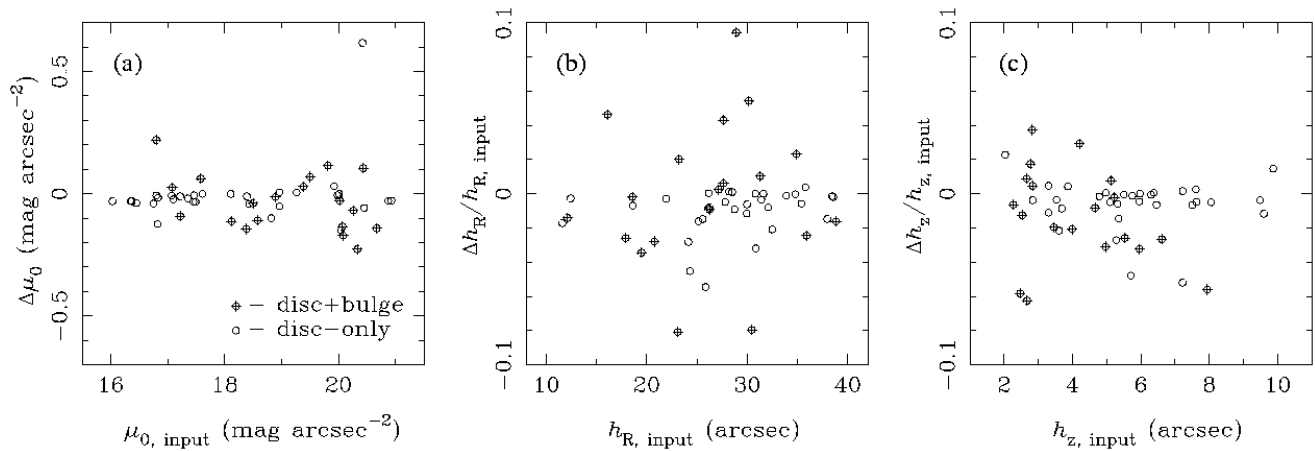}}
	\caption{The recovery of the disc parameters in tests of the
	2D fitting method on artificial images (see text). (a) --
	Edge-on central surface brightness, (b) -- scalelength and (c)
	-- scaleheight.}
\label{fig:tests}
\end{figure*}

\subsection{The approach}
\label{sec:sb_approach}

Since the studies of van der Kruit \& Searle (KS1--3), several
techniques have been developed to model the global structure of
stellar discs in edge-on spiral galaxies \citep{KB87,SG89,X97}. The
application of these techniques to the region near the galactic planes
was found to be problematic, mainly because of three reasons;
(1) the presence of large amounts of dust, (2) the presence of
additional components, e.g. in the form of spiral structure or a young
thin disc, and (3) the limited accuracy, of about 1--2 degrees, with
which the inclination can be determined. 

Photometric studies of large nearby spirals which include a radiative
transfer description, have pointed out that dust is more confined
to the plane than the disc light; ${\it h}_{z,{\rm dust}}\sim
0.5\, {\it h}_{z,{\rm stars}}$ \citep{KB87,X97,X99}. Considering the
relatively small velocity dispersions of young stars
\citep[e.g.][]{W77}, the same is true for young stellar
populations. Hence, if we assume that the range in $z/h_{\rm z}$
affected by dust extinction and young populations is similar among
galactic discs, it becomes possible to mask out that region in a
systematic way. Indeed, the vertical ${\it I - K}$ colour profiles of
the galaxies in our sample, which trace the amount of dust extinction,
show that in all cases the region $|z| > 1.5\ h_{\rm z}$ is only
marginally affected by dust extinction \citep{GPK97}. Therefore, to
solely study the old stellar population, we choose to mask out the
region corresponding to $|z| \leq 1.5\ h_{\rm z}$. Since the
differences among the various functionalities proposed for the
vertical light distribution are small outside this mask \citep[][his
fig.~1]{K88}, the choice for the vertical distribution becomes
unimportant. Still, we will further assess the presence of dust
extinction outside the mask by (1) testing the 2D method on `dusty'
artificial images (Sect.~\ref{sec:sb_tests}) and (2) repeating the 2D
fits in the {\it B} band to make a comparison with the {\it I}-band
results (Sect.~\ref{sec:sb_results}).

The 2D decomposition uses the non-linear least-squares technique
developed by \citet{Ma63}. The fit of
Eqs.~(\ref{eqn:disc}--\ref{eqn:bulge_dev}) is done in the linear  
regime, to the masked version of the sky-subtracted image (the data are
not binned as  in many previous methods, e.g. KS1--3,
\citealt[][]{SG89}). To ensure that the fitting algorithm minimizes
the relative difference between our model and the data, each pixel is
weighted inversely proportional to the value of the model surface
brightness distribution of the disc at that
pixel \citep[cf.][]{J96}. As a consequence of this weighting, regions
of the lowest signal-to-noise carry a large weight in the fit. To
prevent this side effect, the data outside the region defined by 
the $1\sigma$ contour of the model image are excluded from the fit (the
$1\sigma$ level is the standard deviation of the background). By using
this criterion on the model instead of the data, there is no
discrimination in favour of positive noise peaks. Since the adopted
weighting scheme and the fitted region depend on the initial estimate
of the model, an iterative application of the algorithm is required
(see Sect.~\ref{sec:sb_tests}).

The light distribution of galactic discs may have a gradual or sharp
radial steepening in the outer parts, indicating a possible
truncation (e.g. KS1--3). If present, such a region will be excluded
from the 2D fit by using an outer fitting boundary. This boundary will
be assigned by eye, at a projected radius within the radius at
which the radial light profile starts to deviate from that expected
for an exponential disc. The error introduced by the subjectiveness
of this procedure is estimated by performing several 2D fits in which
the fitting boundary is varied by 10--20 percent.

\subsection{Tests on artificial images}
\label{sec:sb_tests}
\normalsize

In order to assess the reliability of the method and the effectiveness
of the mask used at $|z| \leq 1.5\ h_{\rm z}$, a series of tests was
performed on artificial images. First, we created fifty artificial
images using Eqs.~(\ref{eqn:disc}) and (\ref{eqn:bulge_exp}). For the
input parameters, random values were drawn from a uniform distribution
covering the following ranges:

\tabcolsep=1.2mm
\begin{flushleft}
\begin{tabular}{ll}
\emph{Disc} & \emph{Bulge}\\ \hline
16 $<$ $\mu_{0}$ $<$ 21 mag arcsec$^{-2}$ & 17 $<$ $\mu_{\rm e}$ $<$
22 mag arcsec$^{-2}$\\
10 $<$ $h_{\rm R}$ $<$ 40 arcsec & 0.2 $<$ $r_{\rm e}$ $<$
10 arcsec \\
2 $<$ $h_{\rm z}$ $<$ 10 arcsec & 0.5 $<$ $q$ $<$ 1.0\\
\end{tabular}
\end{flushleft}

\noindent
where $\mu_{0}$ and $\mu_{e}$ are used to denote $\Sigma_{0}$ and
$\Sigma_{e}$ in logarithmic units. To ensure the absence of artificial
galaxies resembling bulge-dominated systems, the bulge-to-disc
luminosity ratio was required to be less than two. Then, in order for
the images to resemble the {\it I}-band observations, a sky background
corresponding to $\mu = 20$ mag arcsec$^{-2}$ was added. The use of
a different value for the background in each image is unnecessary,
because sufficient contrast with the background is already provided by
the variation in $\mu_{0}$ and $\mu_{e}$. The effect of seeing was
mimicked by convolving the images with a circular Gaussian of 1.5
arcsec FWHM, which is typical for the observations. Finally, Gaussian
read-out noise and Poisson noise were added and the background was
subtracted.

For each artificial galaxy, initial estimates of the free parameters
were generated by randomly offsetting the input parameters up to $\pm
25$ percent. The region corresponding to $|z| \leq 1.5\ h_{\rm z}$,
was masked out. Then, the 2D fits were performed, while discarding
data outside the $1\sigma$ contour of the model and redefining the
mask and the pixel weights in each iteration. Generally, convergence
was achieved within three iterations. In many of these 2D fits the
bulge parameters did not converge. This is mainly due the use of the
mask, causing the bulge contribution in the fitted region to be
too small (less than 10 percent). However, for the same reason, the 2D
fits could be safely repeated by fitting a disc-only model
(cf. Fig.~\ref{fig:tests}).

Figure~\ref{fig:tests} displays the test results for the disc
parameters. It shows that the scale parameters are always recovered to
within 10 percent, the standard deviation being about 2 percent for
each parameter. As expected, the accuracy of the recovery decreases
with decreasing central surface brightness. For example, the three
outliers of Fig.~\ref{fig:tests}b are all discs of low central surface
brightness, $\mu_{0} > 20$ mag arcsec$^{-2}$. The amplitude of the
deviations shown in Fig.~\ref{fig:tests} is mainly caused by the
application of the mask; if no mask is applied the errors are less
than 1 percent. Note that the application of the mask does not
introduce any systematic error.

We repeated the tests after incorporating a strong dust lane. For the
distribution of the dust, we chose a double exponential with $h_{\rm
R, dust} = h_{\rm R, stars}$ and $h_{\rm z, dust} = 0.5\, h_{\rm z,
stars}$ neglecting scattering \citep[see][their
eq.~(14)]{KB87}. Of course, in reality, dust shows clumpy and
filamentary structures on small scales. However, the global nature of
the fitting routine and the edge-on view largely average these out
\citep{KTGW98}, so that the exponential distribution is a good global
approximation. Our neglect of scattering means that the effect of dust
is slightly overestimated, since scattering would reduce the
extinction. To resemble a strong dustlane, we chose a face-on central
optical depth $\tau_{I}(0) = 1.0$ (cf. \citealt{X99}).

\begin{table*}
\caption[]{{\bf Results of the 2D least-squares fits in the {\it I} band for the
disc component}\\
Columns: (1) Name (ESO-LV catalogue); (2) Hubble type, taken from the
Lyon/Meudon Extragalactic Database (LEDA). These types were assigned
as outlined in \citet{VVC}; (3) Heliocentric velocity corrected for
Virgo-centric flow (LEDA); (4) {\it I}-band absolute magnitude (de
Grijs 1998), calculated using column (3) and corrected for Galactic
extinction; (5) Maximum rotational velocity from LEDA, obtained from a
global H\,{\sc i} profile and corrected for turbulent motions. For the
four galaxies, for which $v_{\rm max}$ is not directly available
(marked by an asterisk), its value was calculated from a least
squares fit to the {\it I}-band Tully-Fisher relation; (6) Neutral
hydrogen mass, calculated using the H\,{\sc i} flux density from LEDA
and column (3); (7) and (8) Edge-on central surface brightness and
error; (9) and (10) Scalelength and error, in arcsec; (11) Scalelength
in kpc; (12) and (13) Scaleheight and error, in arcsec; (14)
Scaleheight in kpc; (15) and (16) The ratio of scalelength to
scaleheight, i.e. the disc flattening, and error}

\tabcolsep=1.4mm
\begin{tabular}{clrrrrrccrrrcccrc}
\hline
Galaxy & type & \multicolumn{1}{c}{{\it v}${\rm _{vir}}$} &
\multicolumn{1}{c}{{\it M}${\rm _{I}^{0}}$} & \multicolumn{1}{c}{{\it
v}$_{\rm max}$} & \multicolumn{1}{c}{{\it M}${\rm _{H\,{\sc i}}}$} &
$\mu_{0,{\rm I}}$ & $\pm$ && $h_{\rm R}$ & $\pm$ & 
\multicolumn{1}{c}{$h_{\rm R}$} & $h_{\rm z}$ & $\pm$ & $h_{\rm z}$ &
$h_{\rm R}/h_{\rm z}$ & $\pm$\\
\cline{7-8}\cline{10-11}\cline{13-14}\cline{16-17}
\noalign{\vspace{2pt}}
&& \multicolumn{1}{c}{(km s$^{-1}$)} & \multicolumn{1}{c}{(mag)} &
\multicolumn{1}{c}{(km s$^{-1}$)} & \multicolumn{1}{c}{(10$^{9}$  
M$_{\sun}$)} & \multicolumn{2}{l}{(mag arcsec$^{-2}$)} &&
\multicolumn{2}{l}{(arcsec)} & (kpc) &
\multicolumn{2}{l}{(arcsec)} & (kpc)\\
(1) & (2) & (3) & (4) & (5) & (6) & (7) & (8) && (9) & (10) &
\multicolumn{1}{c}{(11)} & (12) & (13) & (14) & (15) & (16)\\
\hline
026-G06 & Sc  & 2496 & --19.54 &  99 & 1.38 & 20.4 & 0.1 && 26.0 & 1.5
& 4.19 & 3.7 & 0.3 & 0.60 & 7.0 & 0.7\\
033-G22 & Sc  & 4090 & --19.81 & 113 & --& 19.8 & 0.3 && 18.3 & 5.0 &
4.84 & 2.0 & 0.2 & 0.53 & 9.2 & 2.7\\
041-G09 & Sc  & 4174 & --21.75 & 182 & --& 19.8 & 0.2 && 34.6 & 3.0 &
9.34 & 3.9 & 0.3 & 1.05 & 8.9 & 1.0\\
138-G14 & SBcd& 1508 & --17.73 &  97 & 5.20 & 20.6 & 0.2 && 41.7 & 3.5
& 4.06 & 6.0 & 0.6 & 0.58 & 7.0 & 0.9\\
141-G27 & Sc  & 1646 & --19.08 &  79 & 1.87 & 20.3 & 0.2 && 40.4 & 4.0
& 4.30 & 4.1 & 0.4 & 0.44 & 9.9 & 1.4\\
142-G24 & SBc & 1739 & --19.62 & 112 & 3.10 & 19.7 & 0.2 && 36.8 & 2.7
& 4.14 & 3.8 & 0.2 & 0.43 & 9.7 & 0.9\\
157-G18 & Sc  & 1103 & --18.55 &  89 & 0.87 & 19.8 & 0.2 && 36.5 & 2.2
& 2.60 & 4.8 & 0.3 & 0.34 & 7.6 & 0.7\\
201-G22 & Sbc & 3819 & --20.43 & 155 & 8.52 & 19.3 & 0.2 && 24.3 & 2.5
& 6.00 & 2.9 & 0.4 & 0.72 & 8.4 & 1.4\\
202-G35 & Sbc & 1618 & --19.88 & 123 & 1.54 & 18.3 & 0.2 && 22.8 & 2.3
& 2.38 & 3.6 & 0.4 & 0.38 & 6.3 & 1.0\\
240-G11 & Sb  & 2656 & --21.65 & 260 & 19.61 & 18.9 & 0.4 && 49.0 &
9.0 & 8.41 & 3.5 & 0.8 & 0.60 & 14.0 & 4.1\\
263-G15 & Sc  & 2287 & --21.68 & 155 & 4.34 & 18.6 & 0.2 && 31.2 & 1.6
& 4.61 & 2.9 & 0.2 & 0.43 & 10.8 & 0.9\\
263-G18 & Sbc & 3704 & --22.00 & $^{*}$233 & -- & 19.3 & 0.2 && 27.7 & 2.6
& 6.63 & 3.2 & 0.5 & 0.77 & 8.7 & 1.6\\ 
269-G15 & Sc  & 3191 & --21.16 & 155 & 5.14 & 19.1 & 0.3 && 29.0 & 5.2
& 5.98 & 4.2 & 0.6 & 0.87 & 6.9 & 1.6\\
288-G25 & Sbc & 2333 & --20.84 & $^{*}$164 & -- & 17.6 & 0.4 && 18.4 & 4.0
& 2.77 & 2.5 & 0.4 & 0.38 & 7.4 & 2.0\\
315-G20 & Sbc & 4615 & --21.33 & $^{*}$190 & -- & 19.9 & 0.3 && 22.9 & 5.0
& 6.83 & 2.7 & 0.6 & 0.81 & 8.5 & 2.6\\
321-G10 & Sab & 2965 & --21.06 & 152 & 1.66 & 19.1 & 0.2 && 20.5 & 4.1
& 3.92 & 2.9 & 0.3 & 0.56 & 7.1 & 1.6\\
322-G87 & Sbc & 3464 & --21.41 & $^{*}$195 & -- & 18.1 & 0.4 && 16.8 & 3.5
& 3.76 & 2.6 & 0.4 & 0.58 & 6.5 & 1.7\\
340-G08 & Sc  & 2714 & --19.08 & 106 & 2.52 & 19.1 & 0.1 && 11.5 & 2.8
& 2.02 & 1.4 & 0.2 & 0.25 & 8.2 & 2.3\\
340-G09 & SBc & 2477 & --19.04 &  86 & 2.01 & 20.0 & 0.3 && 22.5 & 3.5
& 3.60 & 3.5 & 0.5 & 0.56 & 6.4 & 1.4\\
416-G25 & Sab & 4812 & --21.53 & 204 & 9.54 & 20.0 & 0.1 && 22.7 & 2.0
& 7.06 & 3.8 & 0.4 & 1.18 & 6.0 & 0.8\\
435-G14 & Sbc & 2482 & --20.11 & 124 & 2.68 & 18.3 & 0.4 && 17.6 & 3.4
& 2.82 & 3.3 & 0.3 & 0.53 & 5.3 & 1.1\\
435-G25 & Sc  & 2290 & --21.46 & 231 & 11.11 & 20.0 & 0.4 && 97.0 &
19.0& 14.36 & 5.0 & 0.3 & 0.74 & 19.4 & 4.0\\
435-G50 & Sc  & 2530 & --18.42 &  77 & 1.79 & 20.1 & 0.1 && 13.2 & 1.5
& 2.16 & 1.7 & 0.3 & 0.28 & 7.8 & 1.6\\
437-G62 & Sa  & 2860 & --22.30 & 209 & 0.71 & 18.9 & 0.5 && 35.8 & 3.7
& 6.62 & 6.6 & 0.5 & 1.22 & 5.4 & 0.7\\
446-G18 & Sb  & 4661 & --21.34 & 189 & 6.08 & 19.1 & 0.1 && 25.8 & 3.0
& 7.77 & 2.0 & 0.1 & 0.60 & 12.9 & 1.6\\
446-G44 & Sc  & 2636 & --20.33 & 149 & 2.81 & 19.0 & 0.2 && 31.1 & 2.1
& 5.30 & 2.8 & 0.2 & 0.48 & 11.1 & 1.1\\
460-G31 & Sc  & 5743 & --22.20 & 226 & 17.50 & 19.5 & 0.3 && 28.1 &
4.6 & 10.43 & 2.4 & 0.4 & 0.89 & 11.7 & 2.7\\
487-G02 & Sb  & 1558 & --20.17 & 162 & 0.29 & 18.4 & 0.3 && 24.3 & 1.9
& 2.45 & 3.8 & 0.4 & 0.38 & 6.4 & 0.8\\
506-G02 & Sbc & 3830 & --21.26 & 179 & 17.25 & 19.0 & 0.1 && 19.4 &
4.0 & 4.80 & 2.8 & 0.4 & 0.69 & 6.9 & 1.7\\
509-G19 & Sbc &10574 & --23.51 & 392 & 13.03 & 19.4 & 0.5 && 24.8 &
4.0 & 16.95 & 2.5 & 0.5 & 1.71 & 9.9 & 2.5\\
531-G22 & Sc  & 3327 & --21.00 & 164 & 4.91 & 18.4 & 0.4 && 19.8 & 4.0
& 4.26 & 3.1 & 0.5 & 0.67 & 6.4 & 1.7\\
555-G36 & SBc & --    & --      & --   & -- & 20.2 & 0.2 && 10.9 & 1.4
& \multicolumn{1}{c}{--}   & 2.8 & 0.3 & -- & 3.9 & 0.7\\
564-G27 & Sc  & 2020 & --19.93 & 152 & 6.75 & 20.7 & 0.2 && 38.0 & 5.1
& 4.96 & 3.5 & 0.4 & 0.46 & 10.9 & 1.9\\
575-G61 & SBc & 1556 & --17.18 &  62 & 0.39 & 20.2 & 0.2 && 16.2 & 1.3
& 1.63 & 2.7 & 0.4 & 0.27 & 6.0 & 1.0\\
\hline
\end{tabular}
\label{tab:dfit2d_disc}
\end{table*}

The effect of the dust on the scale parameters turned out to be
moderate; the $h_{\rm R}$ and $h_{\rm z}$ were found to be 
on average 4 percent and 9 percent larger. The use of a smaller
mask than that adopted here (or no mask at all) would mean that the
effect becomes substantial, causing errors on the scale parameters
larger than 10 percent.

The effect of a moderate stellar warp was investigated by simply
introducing a vertical offset $z_{\rm warp}(R, \phi)$ to the
axisymmetric disc model (Eqn.~\ref{eqn:lumdisc}) for $R > R_{\rm
warp}$:

\begin{eqnarray}
L(R, z, \phi) & = & L(R)\ e^{-|z - z_{\rm warp}(R, \phi)|/h_{\rm z}},\nonumber\\
z_{\rm warp}(R,\phi) & = & w\ [ (R - R_{\rm warp})/h_{\rm R} ]\ \cos(\phi)\ h_{\rm z},
\label{eqn:warpdisc}
\end{eqnarray}

\noindent
where $w$ is the warping rate in units of scaleheight per
scalelength, $R_{\rm warp}$ designates the onset of the warp, and
the line of nodes of the warp is assumed to be straight and along the
line of sight (such that the effect of the warp is the largest). For
any reasonable warping, the effect on the derived scale parameters
turned out to be small. For example, for a stellar warp with $R_{\rm
warp}$ = 3 $h_{\rm R}$ and $w = 0.5$, $h_{\rm R}$ and $h_{\rm z}$ are
respectively under-and over-estimated by only 2 percent. This is
mainly due to the fact that the inner parts of the edge-on disc are
mostly unaffected by the warping.

We also studied the errors introduced by the fixed parameters. This
was done by offsetting these parameters by their typical uncertainty and
repeating the 2D fits. The following fixed parameters were
investigated: sky subtraction -- 1 percent error, seeing -- 0.1\arcsec
error, central position -- 1\arcsec error, position angle -- 1 degree
error, and inclination -- 3 degrees error. For example, for a disc
with $h_{\rm R}/h_{\rm z} = 7$ and an inclination of 87 degrees, our
2D method overestimates the scaleheight by 4 percent. The scalelength
was found to be most sensitive to an error in the sky background,
resulting in an error $\lesssim$ 4 percent. We estimate the total 
error due all systematic effects at $\lesssim$ 6 percent.

Further tests confirmed that the choice for the model used for the
vertical light distribution is rather arbitrary when the region  
corresponding to $|z| \leq 1.5\ h_{\rm z}$ is discarded (as stated in
Sect.~\ref{sec:sb_model}). For example, when instead a vertical
${\rm sech}(z/h_{\rm z})$ distribution is used to create the artificial
images, the scale parameters found by our 2D method differ from the
input values by about 2 percent.

The algorithm does not take into account the effect of seeing on the
disc component. This omission was tested by creating a number of
artificial images, each with a different PSF, and applying the 2D
fitting method. For the unlikely case that the seeing FWHM is
comparable to the scaleheight of the disc, the error made by applying
the method is only 3 percent. Since the scaleheight is much larger
than the seeing FWHM for all galaxies in our sample, one can safely
ignore this effect.

Finally, we investigated the effect of a truncation by applying the
method to truncated artificial galaxies. These were constructed by
including, in Eq.~(\ref{eqn:lumdisc}), $L(R, z) = 0.0$ for $R > R_{\rm
max}$ and numerically integrating along the line of sight. The
difference in edge-on surface brightness between such a truncated disc
and the infinite exponential disc increases with projected radius,
causing the 2D method to underestimate the exponential
scalelength. For galaxies with $R_{\rm max}/h_{\rm R} \geq 4$ this
difference is small. For example, in a galaxy for which $R_{\rm
max}/h_{\rm R}$ = 4, our method underestimates the scalelength by
$\sim$ 6 percent. This error increases slightly for galaxies with
smaller $R_{\rm max}/h_{\rm R}$, reaching $\sim$ 9 percent for $R_{\rm
max}/h_{\rm R}$ = 3.

In summary, we conclude that the uncertainty in the scale parameters
obtained with the 2D method is caused by several effects. In order of
decreasing importance, these are: (1) the presence of a truncation,
the error depending on the value of $R_{\rm max}/h_{\rm R}$, (2)
residual dust extinction (error $\lesssim 9$ percent), (3) inaccurate
sky subtraction (error $\lesssim 4$ percent), and (4) the application of
the mask to $|z| \leq 1.5\ h_{\rm z}$ (error $\sim 2$ percent, see
Fig.~\ref{fig:tests}). For example, for discs with an $R_{\rm
max}/h_{\rm R}$ of 4 (Sect.~\ref{sec:trunc}), the total measurement
error in the scalelength is about 10 percent.

\section{Results}
\label{sec:results}
\subsection{The two dimensional fits}
\label{sec:sb_results}

First, to prepare the images for the 2D fits,  H{\sc ii} regions,
foreground stars and background objects were masked out using a
combination of the {\sc sextractor} algorithm \citep{BA96} and manual
editing in the galaxy's immediate vicinity. The centre and major axis
position angle of the galaxies were determined by fitting ellipses to
the {\it I}-band images using the {\sc galphot} package in {\sc iraf}
\citep*{FIH89b} (see Table~\ref{tab:appendix}, appendix). Then,
several profiles were extracted in the {\it I} band, both parallel and
perpendicular to the major axis. The latter profiles were inspected
for the effects of dust extinction. In most of these profiles, the
turnover due to dust extinction lies below $z = 1.0\ h_{\rm z}$ (see
appendix, and \citealt{GPK97}). However, in 16 galaxies the extinction
clearly extends further out on one side of the galactic plane, e.g. in
ESO 435-G14. This indicates that these galaxies are not exactly
edge-on, and in these cases the affected side was excluded from the 2D
fit.

Fits were made to each {\it I}-band image, taking the values
determined in the 1D analysis \citep{G98} as initial estimates. In 10 
cases, the mask needed to be decreased from $|z| \leq 1.5\ h_{\rm z}$
to $|z| \leq 1.0\ h_{\rm z}$ in order to achieve convergence
(Table~\ref{tab:appendix}, appendix). A bulge-disc decomposition was 
attempted in the 23 systems which clearly show an additional central
component. As the best-fitting bulge model, either exponential
(Eq.~\ref{eqn:bulge_exp}) or r$^{1/4}$ (Eq.~\ref{eqn:bulge_dev}),
the bulge-disc fit with the smallest reduced $\chi^{2}$ was
retained. However, in 14 galaxies the bulge parameters did not
converge. This was not unexpected, since in many galaxies only a small
portion of the bulge light is emitted in the region outside the
mask. Still, two of the 14 systems do show an extended central
component; ESO 321-G10 and ESO 416-G25. In these galaxies this
component has a boxy or peanut-like shape, which probably caused the
fit to diverge. To determine the disc parameters in these 14 cases, we
assigned an inner radial fitting boundary, as in the `marking the
disc' method \citep{F70}, and repeated the 2D fits with a disc
component only.
Tables~\ref{tab:dfit2d_disc} and \ref{tab:dfit2d_bulge} show the
resulting parameters for the entire sample. The errors in the
parameters are {\it not} the formal errors of the least-squares fit;
these were in general less than a few percent and do not take into
account the subjectiveness of the assignment of the radial fitting
boundaries. Instead, the errors were estimated by comparing results
from several similar fits in which the inner and outer radial fitting
boundaries were varied by 10--20 percent. For each galaxy, the 2D fits
are further illustrated in the appendix.

\begin{table}
\caption[]{
{\bf Results of the 2D least-squares fits in the {\it I} band for the
bulge component}\\
Columns: (1) Name (ESO-LV catalogue); (2) Bulge type: e = exponential
Eq.~(\ref{eqn:bulge_exp}), v = de Vaucouleurs
Eq.~(\ref{eqn:bulge_dev}); (3) and (4) Bulge effective surface  
brightness and error; (5) and (6) Bulge effective radius and error;
(7) and (8) Bulge axial ratio and error.}
\tabcolsep=1.4mm
\begin{tabular}{cccccrrccc}
\hline
Galaxy & bulge & $\mu_{\rm e,I}$ & $\pm$ && $r_{\rm e}$ & $\pm$ && $q$
& $\pm$\\
\cline{3-4}\cline{6-7}\cline{9-10}
\noalign{\vspace{2pt}}
&& \multicolumn{2}{l}{(mag arcsec$^{-2}$)} &&
\multicolumn{2}{l}{(arcsec)} &&&\\
(1) & (2) & (3) & (4) && (5) & (6) && (7) & (8)\\
\hline
240-G11 & v & 21.3 & 0.3 && 15.0 & 3.0 && 0.60 & 0.20\\
263-G18 & e & 17.1 & 0.2 &&  2.1 & 0.7 && 0.77 & 0.08\\
315-G20 & v & 19.4 & 0.2 &&  5.7 & 1.3 && 0.56 & 0.09\\
435-G25 & e & 19.2 & 0.1 && 10.4 & 0.3 && 0.50 & 0.03\\
437-G62 & e & 16.0 & 0.5 &&  5.5 & 0.8 && 0.45 & 0.04\\
446-G18 & e & 17.3 & 0.1 &&  4.2 & 0.3 && 0.36 & 0.05\\
460-G31 & v & 19.6 & 0.3 &&  6.8 & 1.5 && 0.45 & 0.14\\
487-G02 & e & 16.3 & 0.2 &&  4.0 & 0.9 && 0.45 & 0.04\\
506-G02 & e & 16.1 & 0.2 &&  2.2 & 0.8 && 0.40 & 0.05\\
\hline
\end{tabular}
\label{tab:dfit2d_bulge}
\end{table}

The 2D fits were repeated in the {\it B} band, using exactly the same
fitting regions to further investigate the effect of dust
extinction. For this, we calculated the scalelength ratios, $h_{\rm
R}^{\rm B}/h_{\rm R}^{\rm I}$. A scalelength ratio is an indirect
measure of the radial colour gradient of the disc. Since dust is known
to be concentrated to the galactic centres and dust extinction
increases towards bluer wavelengths, the presence of dust will result
in a larger $h_{\rm R}^{\rm B}/h_{\rm R}^{\rm I}$. Consequently, when
the observed scalelength ratio is compared to the ratio expected
solely from radial stellar population changes, the effect of dust
extinction can be inferred \citep{PVMF94,G98}. The value of the
scalelength ratio due to population changes is, however, somewhat
uncertain. \citet{PVMF94} argue that the scalelength ratio between the
{\it B} and the {\it I} band due to population changes is in the range
1.05--1.1. From modelling of 6 nearby edge-on galaxies, which includes
a radiative transfer description, \citet{X97,X99} find an average
ratio of 1.25 $\pm$ 0.13 ($1\sigma$). Recently, from a study of 21 Sb
galaxies of various inclination, \citet{C99} derives a ratio of
1.13. For our sample, the average $h_{\rm R}^{\rm B}/h_{\rm R}^{\rm
I}$ is $1.17 \pm 0.18\ (1\sigma)$. This is clearly within the range of
ratios attributed to radial population changes, and smaller than the
ratios observed in less-inclined galaxies \citep{PVMF94}. We conclude
that the dust extinction in the fitted regions is small. This is in
accordance with the results of Sect.~\ref{sec:sb_tests}.

\begin{figure}
	\scalebox{1.0}{\includegraphics{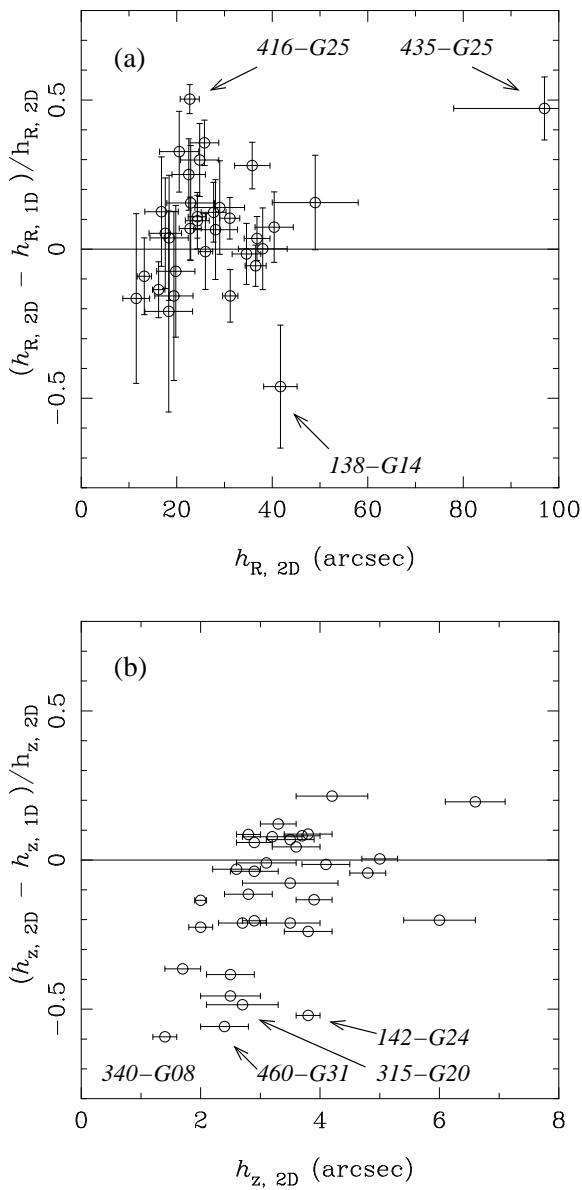}}
	\caption{Comparison of the disc scale parameters obtained
        with the 2D (this paper) and 1D (de Grijs 1998) method. (a) --
        The relative difference between the scalelengths as a function
        of the scalelength determined with the 2D method, (b) -- as
        (a) but for the exponential scaleheight. A few galaxies
        for which there is a large discrepancy are indicated.}
\label{fig:compare}
\end{figure}

\begin{figure}
	\resizebox{8.5cm}{!}{\includegraphics{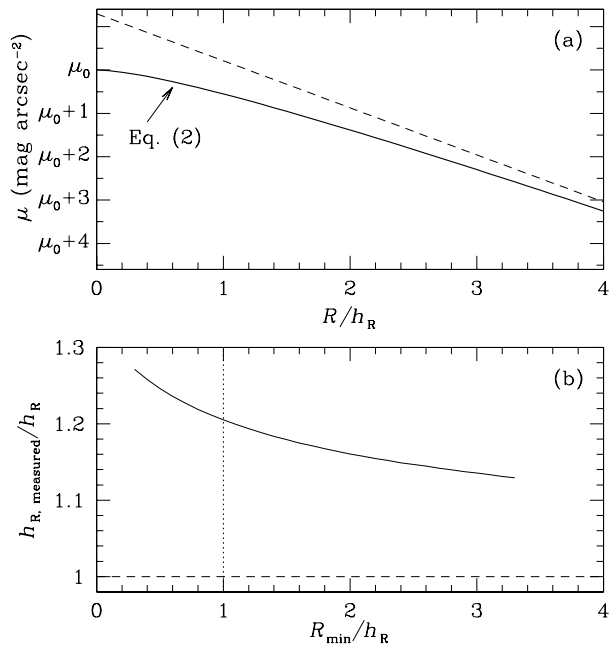}}
	\caption{The systematic error made in the 1D method (de Grijs
	1998) by fitting an exponential to the data instead of
	Eq.~(\ref{eqn:disc}), which leads to scalelengths being
	determined 20 percent too large. (a) -- The radial surface
	brightness profile according to Eq.~(\ref{eqn:disc}) (solid
	line). The exponential profile is shown for comparison (dashed
	line, arbitrary offset). (b) -- The ratio of the measured
	scalelength, obtained in an exponential fit to the region
	between $R = R_{\rm min}$ and $R = 4\, h_{\rm R}$ (de Grijs
	1998), to the actual scalelength as a function of $R_{\rm
	min}$. The vertical dotted line indicates $R_{\rm min}$ =
	$h_{\rm R}$ used by de Grijs (1998).}
\label{fig:1Dwrong}
\end{figure}

In Fig.~\ref{fig:compare}, the disc scale parameters are compared
to those obtained from the 1D method \citep{G98}. For the scalelength,
the differences are significant. The scalelengths determined with the
2D method are on average 13 percent larger. These differences arise
mainly due to the combination of three effects. First, the fitted
regions are different in both methods. According to \citet{KK91},
a difference of about 10 percent is not unusual in this respect. For
example, the 1D method adopts a fixed radial outer fitting boundary
at 4 scalelengths. However, many galaxies show a steepening of the
radial surface brightness in this region, often caused by a truncation
(Sect.~\ref{sec:trunc_results}). This effect causes the scalelength to
be underestimated. Secondly, the bulge contribution is substantial in
early-type spirals, even in the outer parts of the disc, also causing
an underestimation of the scalelength. Thirdly, the 1D method adopts an
exponential as a fitting model, instead of its projection
(Eq.~(\ref{eqn:disc})). This effect leads to an overestimate of the
scalelength by about 20 percent (Fig.~\ref{fig:1Dwrong}).

For the vertical scaleheight, the differences between the two methods
are mostly within 25 percent (Fig.~\ref{fig:compare}b), except for 7
outliers for which the 2D method finds a scaleheight which is smaller
by about 50 percent. These differences are caused by the differences
in the fitting methods; the 2D method uses a two-dimensional fitting
region and a bulge-disc decomposition, whereas in the 1D method the
scaleheight is determined from profiles taken parallel to the major
axis at various projected radii.

Our sample has 5 galaxies in common with the sample of
\citet{PDLS00}. The model used by these authors is similar to that
used here, but the method is different at two important points; (1) --
no mask is applied to the region near the galactic plane and (2) --
whereas with our method the truncation is largely avoided, their
method is based upon the assumption that all discs show an infinitely
sharp truncation. Table~\ref{tab:compare} lists both, the {\it I}-band
scale parameters and truncation radii (Sect.~\ref{sec:trunc_results}),
and the same parameters derived by \citet{PDLS00} in the {\it
r} band. The scale parameters agree well, except for the scalelengths
of two galaxies, ESO 269-G15 and ESO 564-G27. For these, the
determination of \citet{PDLS00} is larger by about 30 percent, much
larger than the difference expected between the {\it I} and the {\it
r}-band from radial population changes alone. We note that for ESO
269-G15, our 2D fit was performed to the eastern side because it is
not perfectly edge-on (see appendix). In ESO 564-G27, additional
extended emission is present at large distances from the plane in the
inner parts, indicating that the use of different fitting regions may
be the cause of the discrepancy.

\begin{table}
\caption[]{{\bf Comparison of the disc parameters with the literature
for five galaxies}\\
Columns: (1) Name (ESO-LV); (2) Band, {\it I} =
{\it I}-band result obtained in this paper, {\it r} = Thuan--Gunn
{\it r}-band result obtained by Pohlen et al. (2000b); (3) and (4)
Scalelength and error; (5) and (6) Scaleheight and error; (7) and (8)
Truncation radius and error (Sect.~\ref{sec:trunc_results}).}
\begin{tabular}{ccccccccrr}
\hline
Galaxy & Band & $h_{\rm R}$ & $\pm$ && $h_{\rm z}$ & $\pm$ && $R_{\rm
max}$ & $\pm$\\
 & & \multicolumn{2}{c}{(arcsec)} && \multicolumn{2}{c}{(arcsec)} &&
\multicolumn{2}{c}{(arcsec)}\\
\cline{3-4}\cline{6-7}\cline{9-10}
\noalign{\vspace{2pt}}
(1) & (2) & (3) & (4) && (5) & (6) && (7) & (8)\\
\hline
269-G15 & {\it I} & 29.0 & 5.2 && 4.2 & 0.6 && 97 & 18\\
            & {\it r} & 40.4 & && 4.0 & && 90&\\
321-G10 & {\it I} & 20.5 & 4.1 && 2.9 & 0.3 && 65 & 6\\
            & {\it r} & 21.6 & && 2.5 & && 65&\\
446-G18 & {\it I} & 25.8 & 3.0 && 2.0 & 0.1 && 84 & 12\\
            & {\it r} & 23.7 & && 2.0 & && 76&\\
446-G44 & {\it I} & 31.1 & 2.1 && 2.8 & 0.2 && 84 & 6\\
            & {\it r} & 29.8 & && 2.9 & && 76&\\
564-G27 & {\it I} & 38.0 & 5.1 && 3.5 & 0.4 && 148 & 11\\
            & {\it r} & 50.9 & && 3.3 & && 141&\\
\hline
\end{tabular}
\label{tab:compare}
\end{table}

\subsection{Fitting the truncation}
\label{sec:trunc_results}

\begin{figure}
\centering
	\scalebox{0.7}{\includegraphics{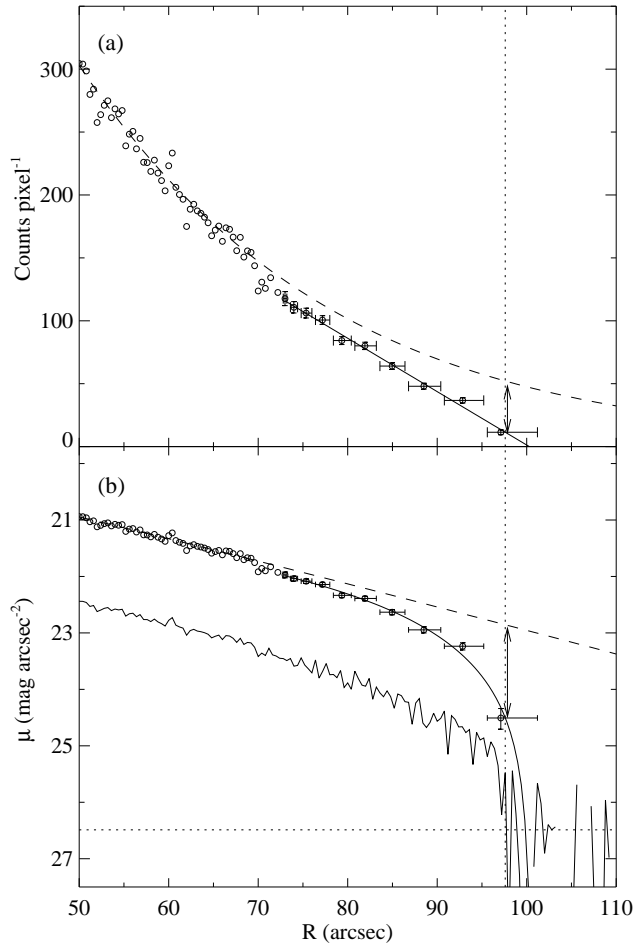}}
	\caption{The edge-fitting method, illustrated with a close-up
	of the averaged profile on one side of an artificial
	galaxy. This artificial galaxy was constructed as in
        Sect.~(3.3) with $R_{\rm max} = 100$ arcsec and $R_{\rm
	max}/h_{\rm R} = 4$. (a) -- Linear surface brightness as a
	function of radius after vertical and radial averaging (open
	circles). The profile expected for an untruncated exponential
	disc (Eq.(2)) is shown by the dashed line. The vertical dotted
	line marks the outermost radius for which data is included in
	the trial fits. In this example, the minimum
	reduced $\chi^{2}$ is reached when including 10 points in the
	fit (after trying 4 points and 7 points); the corresponding
	best-fitting straight line is shown as a solid
	line. Extrapolation of this fit to zero flux yields $R = 100.3
	\pm 3.6$ arcsec. The difference in surface brightness with the
	untruncated disc at the outermost radius is marked by the
	arrow. It corresponds to 3$\sigma$ of the background. (b) --
	as (a) but on a logarithmic scale. In addition, the vertically
	averaged profile is shown, shifted by 1.5 mag
	arcsec$^{-2}$. The horizontal dotted line, shifted by the same
	amount, marks the standard deviation of the background in the 
	vertically averaged profile (see text for further details).}
\label{fig:trunc_illus}
\end{figure}

The truncation, or cut-off, of a stellar disc can be inferred from a
sharp drop in the radial light profile (e.g. KS1--3). The truncation
radius is then obtained by extrapolating the radial light profile
to zero flux. The error made in such an extrapolation is largely
due to Poisson statistics, imperfect sky subtraction and foreground
stars (KS1--3). While truncations are most easily observed in edge-on
systems because of their higher surface brightness, there one views
the result of a complicated line of sight integration across the
truncated disc, which includes the effects of dust extinction and
non-axisymmetric components such as spiral arms. Additionally, the
truncation may not be axisymmetric, such that in an edge-on spiral the
truncation becomes apparent at a different projected radius on either
side \citep{GKW01}. These considerations, as well as the variety of
shapes of the radial light profiles seen in truncated edge-on spirals
\cite[e.g.][]{NJ97,GKW01}, illustrate the difficulties faced in the
construction of a model for the shape of the truncation from studies
of edge-on spirals. For these reasons we did not include a truncation
in the model adopted for the 2D fits (Sect.~\ref{sec:sb_method}).

By applying an edge-fitting method similar to the one introduced by
\citet{GKW01}, it is possible to objectively determine the truncation
radius. The adopted method is as follows. A radial profile is
constructed by first averaging vertically over the range 
$1.0\ h_{\rm z} \leq z \leq 2.0\ h_{\rm z}$ on the least obscured side
of the galactic plane. The choice for this range minimizes the effects
of dust and young populations, while retaining a good signal-to-noise
ratio. To further increase the signal-to-noise in the outer parts the
profile is also averaged radially, by using a semi-logarithmic scheme
\citep{GKW01}. We made sure that this radial averaging does not smooth
out the truncation by testing the edge-fitting method on artificial
images (these were constructed as in Sect.~\ref{sec:sb_tests} and
included the infinitely sharp truncation). Then, the part of the
profile containing flux from the galaxy is defined by clipping the
vertically averaged profile at the 1$\sigma$ level ($\sigma$ being the
standard deviation of the background in the vertically averaged
profile), and using it as a mask (i.e. a method of conditional
transfer). In this way, the outermost projected radius included in the
profile is determined directly by the data quality. Following this, a
series of trial fits of a straight line is performed in linear
surface brightness at the edge of the profile (note that \citet{GKW01}
used an exponential function). In the first trial fit the four
outermost points are included. Then, in each of the subsequent trial
fits the inner fitting boundary is decreased by three points. Finally,
only the fit with the smallest reduced $\chi^{2}$ is retained and
extrapolated to zero flux. The corresponding projected radius is then
the adopted truncation radius, provided it satisfies the criterion
outlined in the following.

An edge-fitting method was already applied, although in a somewhat
different form, to four sample galaxies in {\it B}, {\it V}, and {\it
I} band \citep{GKW01}. There, as in all previous studies, the presence
of a truncation is deduced from a qualitative comparison between
the observed radial profile and the exponential disc model. In our
sample, which is expected to contain many less obvious cases, a
quantitative criterion is needed to determine its presence. To search
for such a criterion, we performed tests of the edge-fitting method on
artificial images. Not surprisingly, it was found that $R_{\rm max}$ is more
easily recovered when the difference between the observed surface
brightness and that expected for an untruncated exponential disc
(Eq.~(\ref{eqn:disc})) is large. In particular, when the difference
between the observed and expected surface brightness at the outermost
radius of the profile is larger than about 2$\sigma$, then $R_{\rm
max}$ could always be recovered ($\sigma$ is the standard deviation of
the background in the original image). When the difference at the
outermost radius is smaller than 2$\sigma$, which is caused by either
a low central surface brightness and/or large $R_{\rm max}/h_{\rm R}$,
then the recovery often fails. In what follows, a profile is said to
be truncated if it satisfies this 2$\sigma$ criterion. The
edge-fitting method is further illustrated in
Fig.~\ref{fig:trunc_illus}.

Of course, when measuring surface brightnesses in the outer parts of
galactic discs, the accuracy of the background subtraction becomes
very important. The main factors which limit this accuracy are large
scale flatfielding errors and the presence of background sources and
foreground stars; our small-scale flatfielding errors are within 0.5
percent \citep{G97} and the CCDs were not affected by fringing. The
background emission in the {\it I}-band images was originally
estimated by fitting a plane to regions far away from the galaxy
\citep{G97}. To check the influence of residual light due to objects
within that region, we repeated these fits after heavily growing of
our object masks. The results show that the earlier fits slightly
overestimated the background, on average by 0.3 times the standard
deviation in the background. Although this oversubtraction is small,
much smaller than the amplitude necessary to produce artificial 
truncations (see below), it would affect the deduced values for
$R_{\rm max}$ and we have therefore decided to apply the edge-fitting
method to the new background-subtracted images. To estimate the
amplitude of remaining variations due to large-scale flatfielding
errors, we inspected the distributions of pixel values in 10x10 sized
boxes at a large number of positions in the sky-subtracted background
images. The standard deviation of the medians of these distributions
was adopted as the error in the background determination.

\begin{table}
\caption[]{{\bf Results of the successful truncation fits in the {\it
I} band}\\
Columns: (1) Name (ESO-LV), an asterisk indicates a galaxy which is
not exactly edge-on and shows signs of spiral structure; (2) Side;
(3) and (4) Truncation radius and error, in arcsec; (5) and (6)
Truncation radius and error, in {\it I}-band scalelengths; (7)
Truncation radius in kpc; (8) Difference between the observed surface
brightness and that of the untruncated exponential disc
(Table~\ref{tab:dfit2d_disc}) at the outermost point of the profile,
in units of the standard deviation of the background.}
\begin{tabular}{llrrcccrr}
\hline
Galaxy & Side & $R_{\rm max}$ & $\pm$ && $R_{\rm max}$ & $\pm$ &
$R_{\rm max}$ & Diff.\\

& & \multicolumn{2}{c}{(arcsec)} && \multicolumn{2}{c}{($h_{\rm R}$)} &
\multicolumn{1}{r}{(kpc)} & \multicolumn{1}{r}{($\sigma$)}\\
\cline{3-4}\cline{6-7}
\noalign{\vspace{2pt}}
\multicolumn{1}{c}{(1)}&(2)&(3)&(4)&&(5)&(6)&(7)&(8)\\
\hline
026-G06 & W & 85 & 4 && 3.3&0.3& 14 &  3\\
041-G09$^{*}$ & SE&125 & 4 && 3.6&0.3& 34 &  8\\
141-G27 & SE&104 & 2 && 2.6&0.3& 11 &  7\\
        & NW&110 &10 && 2.8&0.4& 12 &  8\\
142-G24$^{*}$ & S &130 & 5 && 3.5&0.3& 15 & 13\\
        & N &126 & 9 && 3.4&0.3& 14 & 13\\
157-G18 & NE&127 &18 && 3.5&0.5&  9 &  5\\
        & SW&143 &18 && 3.9&0.5& 10 &  4\\
201-G22 & NE&102 & 5 && 4.2&0.5& 25 &  4\\
        & SW& 96 & 7 && 3.9&0.5& 24 &  6\\
202-G35$^{*}$ & NW& 95 & 6 && 4.1&0.5& 10 &  6\\
240-G11$^{*}$ & SE&210 &14 && 4.3&0.8& 36 & 11\\
269-G15$^{*}$ & N & 97 &18 && 3.4&0.9& 20 &  5\\
288-G25$^{*}$ & NE& 91 &19 && 5.0&1.5& 14 &  7\\
315-G20 & NE& 71 & 4 && 3.1&0.7& 21 &  6\\
321-G10 & NE& 65 & 6 && 3.2&0.7& 12 &  5\\
416-G25 & NE& 74 & 4 && 3.3&0.4& 23 &  5\\
        & SW& 74 & 3 && 3.3&0.3& 23 &  5\\
435-G14$^{*}$ & SW& 79 &12 && 4.5&1.1& 13 &  6\\
446-G18$^{*}$ & S & 84 &12 && 3.3&0.6& 25 & 13\\
446-G44 & W & 79 & 6 && 2.6&0.3& 14 & 26\\
        & E & 88 & 6 && 2.8&0.3& 15 & 25\\
460-G31$^{*}$ & W &113 &24 && 4.0&1.0& 42 &  7\\
487-G02 & NE&113 &11 && 4.6&0.6& 11 &  9\\
        & SW&110 & 5 && 4.5&0.4& 11 & 10\\
509-G19$^{*}$ & SW& 90 &12 && 3.6&0.8& 62 &  8\\
        & NE& 91 & 9 && 3.7&0.7& 62 &  9\\
564-G27$^{*}$ & N &148 &11 && 3.9&0.6& 19 &  8\\
\hline
\end{tabular}
\label{tab:truncfit}
\end{table}

\begin{table}
\caption[]{{\bf Lower limits to the truncation radii for the
unsuccessful truncation fits}\\
Columns: (1) Name (ESO-LV); (2), (3) and (4) Lower limit to the
truncation radius in arcsec, {\it I}-band scalelengths and kpc
respectively}
\begin{tabular}{lrrr}
\hline
Galaxy & \multicolumn{3}{c}{$(R_{\rm max})_{\rm min}$}\\
& (arcsec) & ($h_{\rm R}$) & (kpc)\\
\noalign{\vspace{2pt}}
\multicolumn{1}{c}{(1)}&(2)&(3)&(4)\\
\hline
033-G22 &  80 & 4.4 & 21\\
138-G14 & 130 & 3.1 & 13\\
263-G15 & 113 & 3.6 & 17\\
263-G18 &  80 & 2.9 & 19\\
322-G87 &  60 & 3.6 & 13\\
340-G08 &  70 & 6.1 & 12\\
340-G09 &  70 & 3.1 & 11\\
435-G25 & 200 & 2.1 & 30\\
435-G50 &  60 & 4.5 & 10\\
437-G62 & 115 & 3.2 & 21\\
506-G02 &  65 & 3.4 & 16\\
531-G22 &  85 & 4.3 & 18\\
555-G36 &  65 & 5.0 & --\\
575-G61 &  55 & 3.4 &  6\\
\hline
\end{tabular}
\label{tab:minrmax}
\end{table}

The results of the application of the edge-fitting method to our
sample are shown in Table~\ref{tab:truncfit} (and the appendix). To
calculate the error on $R_{\rm max}$ two additional images were
constructed for each galaxy by adding/subtracting the error in the
background determination. The fits were repeated and the error on
$R_{\rm max}$ was calculated by quadratically adding half the
difference between the truncation radii found from the two modified
images to the formal error of the original fit. For the galaxies for
which no truncation could be determined, Table~\ref{tab:minrmax} shows
a lower limit for $R_{\rm max}$.

To summarize, a truncation is found in 28 of the 68 profiles (counting
each side of each galaxy separately). At the outermost point in these
profiles, the differences between the exponential disc model and the
data are larger than $\sim 3\sigma$, clearly satisfying our
criterion. These differences are also much larger than the error in 
the background determination, i.e. the truncations are not produced by
over-subtracting the sky background (Table~\ref{tab:truncfit} and the
appendix). Of the other 40 profiles, about half had to be discarded
because of the presence of foreground stars, background objects or a
limiting field of view. In the remaining profiles, the difference
between the data and the untruncated exponential disc at the outermost
point is less than 2$\sigma$. The 28 truncations are found in 20
galaxies, 8 of which show a truncation on both sides. In all of these
8 cases the truncation appears symmetric; any asymmetries are less
than 15 percent. Of the 12 galaxies in which a single truncation is
seen, there are only 2 for which the profile at the other side of the
galaxy could be studied as well; ESO 435-G14 and ESO 460-G31. Looking
directly at the images of these, we note that their discs are 
not perfectly edge-on and show a signature of strong spiral arms which
may be causing the single-sided truncation \citep[cf.][]{GKW01}.
Galaxies which are not exactly edge-on and show signs of spiral
structure are marked with an asterisk in Table~\ref{tab:truncfit}. As
an alternative possibility, the single truncations may be caused by a
non-axisymmetric or lopsided stellar disc in which the truncation
occurs at a smaller projected radius (= higher surface brightness) on
one side.

To assess the effect of dust extinction, the fits were repeated using
the region below one scaleheight. The resulting $R_{\rm max}$ are
consistent with the values of Table~\ref{tab:truncfit}, indicating
that dust extinction near the edge of the disc is not very
important. Observations of dust tracers such as CO emission and
mm/submm continuum in the outer parts of spiral galaxies confirm this
picture \citep{CB97,NGGZW96}.

Other effects which may cause an apparent truncation are warping or
flaring of the stellar disc. Although our sample contains no spirals
which are strongly warped (Sect.~\ref{sec:sample}), artificial images
of warped discs (Sect.~\ref{sec:sb_tests}) show that a moderate
warping can still introduce a break in the radial profile similar to
the signature of a truncation. More importantly, away from the major
axis the radial profiles show a strong asymmetry; 
a break at a small galactocentric radius on one side and an upturn
followed by a break at a large galactocentric radius on the
other. This causes a large difference between the apparent truncation
radii of both sides. While similar upturns in a few of our sample
galaxies may be a sign of warping (or the effect of spiral arms), the
galaxies having a double-sided truncation show no evidence for such a
strong asymmetry. To address disc flaring we also investigated
artificial images of flared discs viewed edge-on. These which were
constructed as in Sect.~\ref{sec:sb_tests}, but included a scaleheight
increasing linearly with radius. It turned out that a very strong and
localized thickening in the outer parts is required to produce a
feature resembling a radial truncation (a rate larger than 1 $h_{\rm
z}$ per $h_{\rm R}$). For our sample galaxies, and late-type spirals
in general, such thickening is not observed \citep{GP97,MBH94,FMHB99}.
We conclude that warping and flaring are very unlikely to have caused
the truncations.

Again, we can compare our results to those obtained by \citet{PDLS00},
who estimated the truncation radius by visually comparing radial
profiles to a truncated model (Table~\ref{tab:compare}). For these 5
galaxies the results for the truncation radius are consistent. Thus,
the truncation is seen in different data-sets using different methods,
indicating that the truncations are real. This is further emphasized
by the tight relation observed between $h_{\rm R}$ and $R_{\rm max}$
(Sect.~\ref{sec:trunc}).

Finally, we would like to emphasize that the fraction of galaxies in
which a truncation is found, about 60 percent, is a lower
limit. Clearly, the detection of a truncation depends on the edge-on
central surface brightness, on the $R_{\rm max}/h_{\rm R}$ of the disc,
on the limiting magnitude reached by the observation and, to a lesser
extent, on the shape of the truncation. Since these elements are
different in each image, the selection effect is different in
each case.

\section{Discussion}
\label{sec:discussion}

\subsection{The disc flattening}
\label{sec:flat}

\begin{figure*}
\centering
	\scalebox{0.74}{\includegraphics{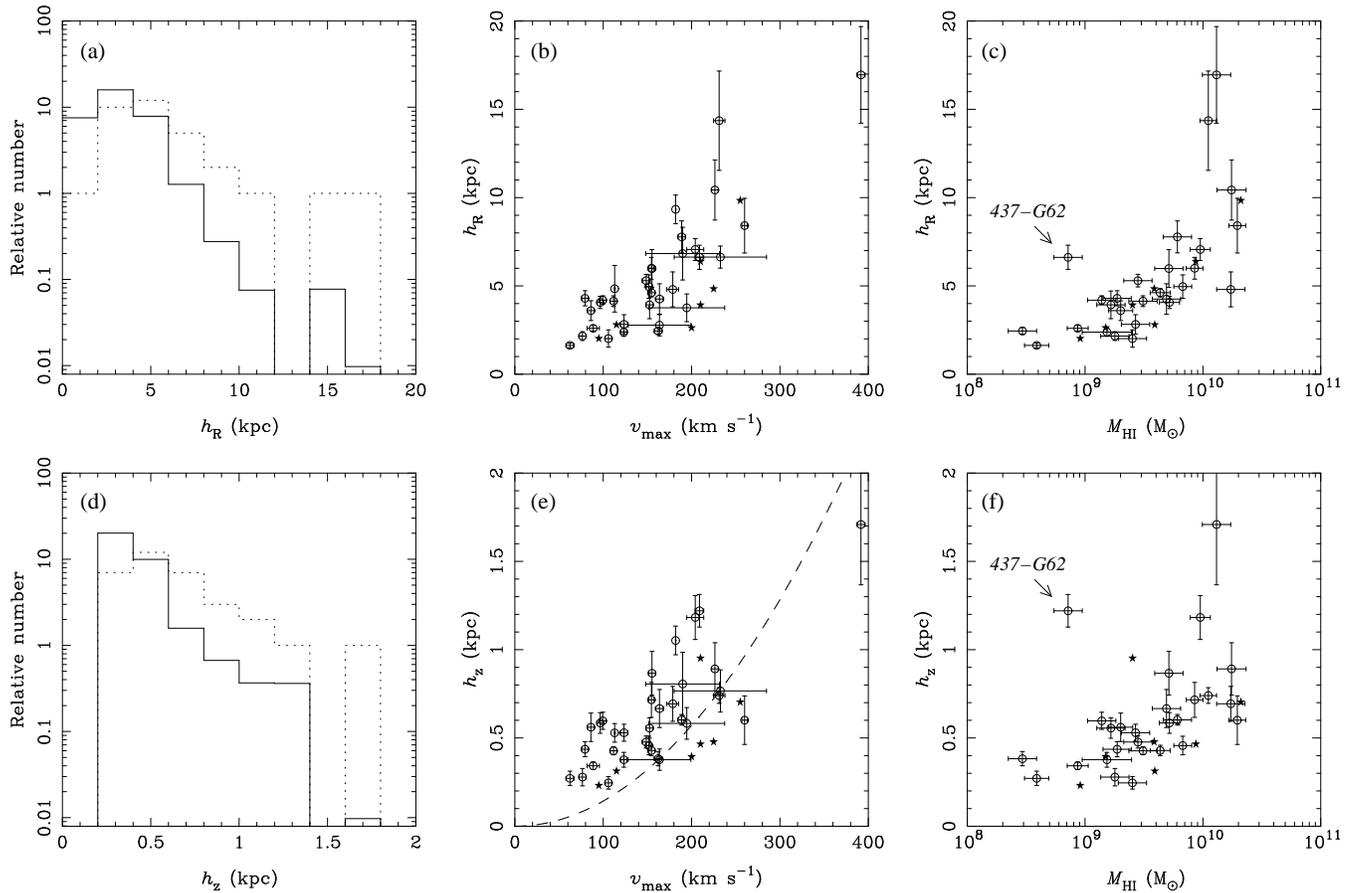}}
	\caption{The disc scale parameters in physical units; (a) --
	The scalelength distribution before (dotted histogram) and
	after (solid histogram) the volume correction (see text), (b)
	-- Scalelength as a function of maximum rotational velocity,
	(c) -- Scalelength as a function of total H\,{\sc i} mass,
	(d)\&(f) -- As (a)\&(c) but for the scaleheight. The dashed
	line in panel (e) is the relation expected for a vertically
	exponential disc with face-on central surface brightness
	$\mu_{B} = 21.65$ mag arcsec$^{-2}$ and constant mass-to-light
	ratio $M/L_{\rm B}$ = 2 ${\rm (M/L)}_{\sun}$ which obeys the
	vertical Jeans equation and Eq.~(\ref{eqn:bottema}). In panels
	(b), (c), (e), and (f) the pentagrams indicate the galaxies
	from KS1--3.}
\label{fig:scale}
\end{figure*}

We further investigate the disc scale parameters obtained in
Sect.~\ref{sec:sb_results}. In Figs.~\ref{fig:scale}a$\,$\&$\,$d 
the distributions of the {\it I}-band scalelengths and
scaleheights are plotted. The corrected distributions were obtained by
applying a volume correction to the diameter-limited distributions
\citep{K87,D90,JIII96} and normalizing to the number of galaxies in the
sample. The diameter selection limit implies that at small physical
sizes the sampled volumes are too small to contain any edge-on spirals
(from our sample the density of spirals with inclination $\geq$ 87
$^{\circ}$ is only about 2 x 10$^{-3}$ Mpc$^{-3}$). This selection effect
causes the low number of small spirals seen in the scalelength
distribution. Small scalelength spirals are known to dominate the
population of spirals \citep{K87,JIII96}. Recognizing the fact that
these small spirals are underrepresented, the shape of the
volume-corrected distribution is consistent with the results of
\citet{K87} and \citet{JIII96}. Interestingly, the distribution of
scaleheight shows a similar shape, including a lack of small
scaleheight galaxies (Fig.~\ref{fig:scale}d). This lack is also due to
the selection against intrinsically small galaxies because scaleheight
and scalelength are observationally related (see below). Still, the
smaller scaleheight discs are the most numerous; the volume-corrected
distribution of scaleheights implies that about 90 percent of the
spirals has $h_{\rm z} < 0.6$ kpc.

\begin{figure*}
\centering
	\scalebox{0.80}{\includegraphics{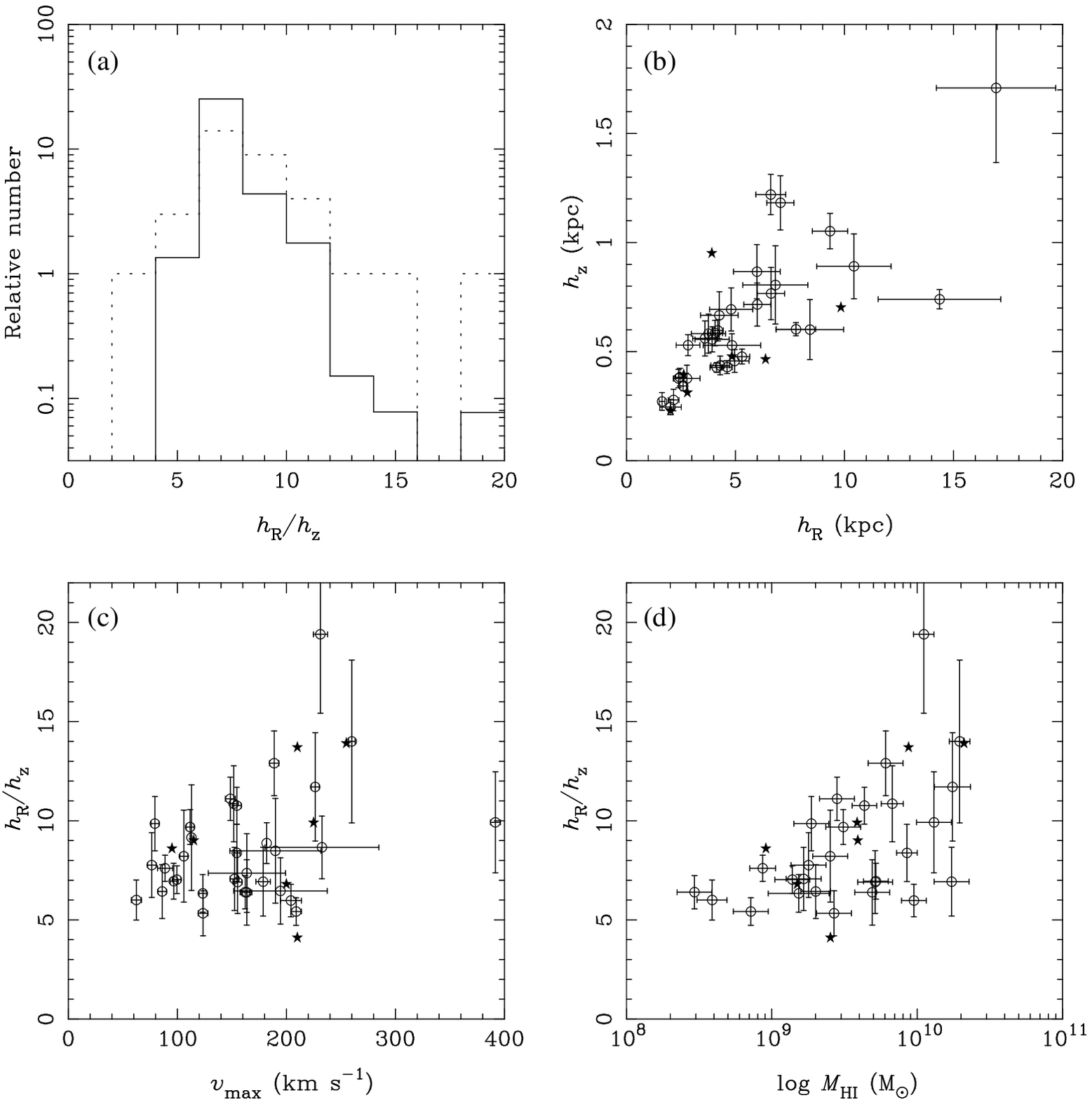}}
	\caption{The disc flattening; (a) -- The flattening distribution
	before (dotted histogram) and after (solid histogram) volume
	correction, as described in the text, (b) -- Scalelength
	versus scaleheight, (c) -- Flattening as a function of
	maximum rotational velocity, (d) -- Flattening as a function of
	total H\,{\sc i} mass. In panels (b), (c) and (d) the
	pentagrams indicate the galaxies from KS1--3.}
\label{fig:flat}
\end{figure*}

For 33 spirals with known maximum rotational velocities
(Table~\ref{tab:dfit2d_disc}), both scale parameters are plotted as a
function of maximum rotational velocity in
Fig.~\ref{fig:scale}b$\,$\&$\,$e. This shows clearly that the disc
scalelength increases with maximum rotational velocity. A Spearman
rank correlation test results in a coefficient of 0.75, corresponding
to a confidence level greater than 99 percent. Since the maximum
rotational velocity is an indicator of total mass, this merely shows
that more massive spirals tend to have larger stellar discs. More
interestingly, the disc scaleheight shows a similar behaviour
(Spearman correlation coefficient 0.74). Hence, more massive spiral
galaxies have both larger {\it and} thicker discs. The most extreme
example of this is ESO 509-G19. This huge spiral has an extremely large
maximum rotational velocity of \mbox{392 km s$^{-1}$} and a disc
scalelength and scaleheight of 17 kpc and 1.7 kpc respectively. A
similar correlation is seen between the scale parameters and the total
H\,{\sc i} mass (Fig.~\ref{fig:scale}c$\,$\&$\,$f), with the notable
exception of ESO 437-G62 (= NGC 3390). This outlier has a remarkably
small amount of H\,{\sc i} and a thick disc, and is classified as an
S0--a in the ESO-LV catalogue. For spirals however, more massive
H\,{\sc i} layers tend to be associated with larger and thicker
stellar discs. We observe no clear trend with morphological type.

Observations of the stellar kinematics in 12 spirals by \citet{B93}
have revealed a related trend; more massive spirals tend to have 
larger stellar velocity dispersions. A simple straight line fit,
adopted from his fig.~6, gives:

\begin{equation}
\sigma_{\rm z}|_{\rm R=0} = (0.29\pm0.10)\ v_{\rm max},
\label{eqn:bottema}
\end{equation} 

\noindent
where $\sigma_{\rm z}|_{\rm R=0}$ is the central vertical velocity
dispersion of the disc stars. This vertical velocity dispersion is
directly related to the scaleheight of the disc mass distribution
through the vertical Jeans equation. When the vertical velocity
dispersion is integrated in the vertical direction \citep[see][]{K88},
one has:

\begin{equation}
\sigma_{\rm z}|_{\rm R=0} = \sqrt{C G \Sigma_{0} h_{\rm z}},
\label{eqn:hydrostat}
\end{equation}

\noindent
where $\Sigma_{0}$ is the central disc surface density and $C$ is a
constant which depends on the exact choice of the vertical density
distribution; for an exponential distribution $C$ = $3\pi/2$, for a 
sech-dependence $C$ = 1.7051 $\pi$ and for the locally isothermal
distribution $C$ = 1. If we assume that the scaleheight of the disc
mass distribution is proportional to that of the old stellar
population, Eq.~(\ref{eqn:hydrostat}) predicts that an increase in the
vertical velocity dispersion is associated with an increase in the
scaleheight of the light of the old population. Furthermore, by
eliminating $\sigma_{z}|_{R=0}$ between Eq.~(\ref{eqn:bottema}) and
(\ref{eqn:hydrostat}), the scaleheight of the disc light is then also
expected to be related to the maximum rotational velocity; $h_{z}
\propto v_{\rm max}^{2}/\sqrt{\Sigma_{0}}$. As an example, we
illustrate this in Fig.~\ref{fig:scale}e (dashed line) for vertically
exponential discs with face-on central surface brightness $\mu_{B} =
21.65$ mag arcsec$^{-2}$ \citep{F70} and constant mass-to-light ratio
$M/L_{\rm B}$ = 2 ${\rm (M/L)}_{\sun}$. While the disc mass-to-light
ratios in and among spiral galaxies are still poorly known, at least
the shape of the predicted relation for a constant mass-to-light ratio
is roughly similar to that of the observations.

In Fig.~\ref{fig:flat} we investigate the ratio of the {\it I}-band
scale parameters, which is a direct measure of the intrinsic
flattening of the disc light. The volume corrected distribution of the
disc flattening (Fig.~\ref{fig:flat}a) is remarkably narrow; about 70
percent of spirals seem to have a flattening between 6 and 8. The
average flattening of the diameter-limited distribution is
$\left<h_{\rm R}/h_{\rm z}\right> = 8.5 \pm 2.9$ (1$\sigma$), while
that of the volume corrected distribution is $\left<h_{\rm R}/h_{\rm
z}\right>=7.3 \pm 2.2$ (1$\sigma$). Figure~\ref{fig:flat}b reveals a
clear correlation between $h_{\rm R}$ and $h_{\rm z}$ (Spearman
correlation coefficient 0.84). Considering the errors, the spread in
the $h_{\rm R}$--$h_{\rm z}$ plane is real, suggesting that the
physical relation between $h_{\rm R}$ and $h_{\rm z}$ cannot be
described by a simple linear functionality expected for discs of
constant flattening. This 
is even more clear in Fig.~\ref{fig:flat}c$\,$\&$\,$d. The range and
maximum of the disc flattening seem to increase with both maximum
rotational velocity and total H\,{\sc i} mass. A similar behaviour is seen
with absolute magnitude $M_{\rm I}^{0}$ (not shown). The disc
flattening shows a similar, but weaker, trend with Hubble type
\citep{G98,SD00}, in accordance with the results on the disc axial
ratio obtained by \citet{Gu92} and \citet{K94}. On the other hand, the
minimum disc flattening seems to be independent of all of these
parameters. At present, there is no clear physical explanation for
these trends.

The disc flattening can be used to estimate the contribution of the
disc to the rotation curve. For a self-gravitating exponential thin
disc, there is a relation between maximum rotational velocity ($v_{\rm
max}^{\rm disc}$) and disc scalelength \citep{F70}:

\begin{equation}
v_{\rm max}^{\rm disc} = 0.88\ \sqrt{\pi G \Sigma_{0} h_{\rm R}}.
\end{equation}

\noindent
When the central surface density is substituted from
Eq.~(\ref{eqn:hydrostat}), a simple relation between the maximum
rotation, vertical velocity dispersion and flattening of the disc is
obtained \citep{B93}:

\begin{equation}
v_{\rm max}^{\rm disc} = (0.69 \pm 0.03)\ \sigma_{\rm z}|_{\rm R=0} \sqrt{\frac{h_{\rm R}}{h_{\rm z}}},
\label{eqn:selfgrav_disc}
\end{equation}

\noindent
where the constant accounts for density laws ranging from an
exponential to a sech-dependence, and $h_{\rm R}/h_{\rm z}$ is the
flattening of the disc mass. This equation states that a flatter disc
with the same vertical stellar velocity dispersion or `temperature'
will be more massive. By eliminating $\sigma_{\rm z}$ between this
equation and the empirical relation Eq.~(\ref{eqn:bottema}), one can
estimate the contribution of the disc to the rotation curve:

\begin{equation}
v_{\rm max}^{\rm disc}/v_{\rm max} = (0.21\pm0.08)\ \sqrt{\frac{h_{\rm R}}{h_{\rm z}}}.
\end{equation}

\begin{figure*}
\centering
	\scalebox{0.80}{\includegraphics{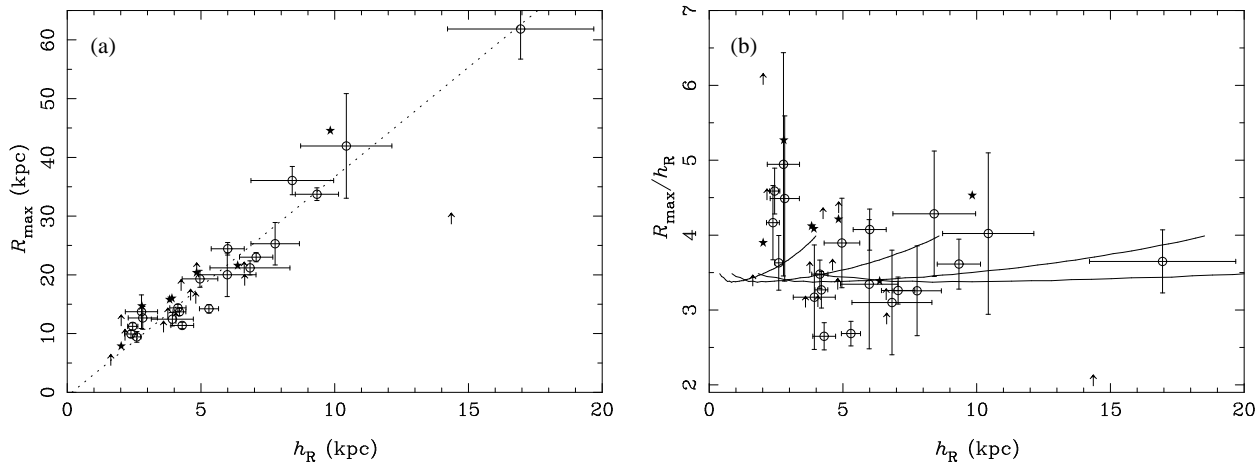}}
	\caption{(a) -- The truncation radius versus disc
	scalelength. The dotted line is the least-squares fit of a
	straight line. The arrows indicate a lower limit for 13
	galaxies for which no $R_{\rm max}$ could be determined. The
	pentagrams indicate the galaxies from KS1--3. (b)
	-- $R_{\rm max}/h_{\rm R}$ as a function of disc
	scalelength. The arrows and pentagrams are as in (a). The
	lines illustrate the prediction of a collapse model for disc
	galaxy formation (see text). Each line connects models of the
	same total mass, but with a varying spin parameter, $\lambda$,
	ranging from 0.01 (at the left end of each line) to 0.28 (at
	the right end of each line). The total mass increases in
	powers of ten, from 10$^{10}$ M$_{\sun}$ (for the line at the
	left) to 10$^{13}$ M$_{\sun}$ (for the line at the right).}
\label{fig:trunc}
\end{figure*}

\noindent
By taking $h_{\rm R}/h_{\rm z}$ = 10 for the average flattening of the
disc mass and using a locally isothermal vertical distribution,
\citet{B93} argued that the disc supplies on average 63 percent 
of the observed maximum rotation, i.e. $v_{\rm max}^{\rm disc}/v_{\rm
max} = 0.63$. Of course, this argument can be used for our sample,
where we measured the flattening of the disc light in the {\it I}
band. Assuming that the flattening of the disc mass equals that of
the disc light from the old stellar population, and using the volume
corrected average of the flattening of $7.3 \pm 2.2$, we find
$\left<v_{\rm max}^{\rm disc}/v_{\rm max}\right> = 0.57\pm0.22$. This
confirms the result of \citet{B93} and suggests that in a typical
spiral galaxy, the actual disc mass-to-light ratio is about one third
of that expected in a maximum disc situation \citep{A85,F92}. However,
the result is not entirely unambiguous. Given the error in the
determination, the value of the disc contribution to the rotation
could be about 80 percent. This would mean the disc can still be close
to maximum, since, to avoid hollow halo cores and allow for a bulge
contribution, a maximum disc has $v_{\rm max}^{\rm disc}/v_{\rm max}
\sim 0.9$. The relatively large error in our determination stems
from the scatter in the empirical relation Eq.~(\ref{eqn:bottema}), a
relation which is still ill-constrained due to the small number of
spirals for which stellar velocity dispersions are known.

\subsection{The disc truncation}
\label{sec:trunc}

The truncation radius is an important global parameter, for which we
need to find an explanation in theories of disc galaxy formation
and/or evolution. In Sect.~\ref{sec:trunc_results}, we showed that the
light distribution of at least 20 out of the 34 galaxies in our sample
is radially truncated. These galaxies display a remarkably tight
relation between the truncation radius and disc scalelength
(Fig.~\ref{fig:trunc}a) (Spearman correlation coefficient 0.95). By
taking the average $R_{\rm max}$ of both sides in case of a
double-sided truncation, we obtain an average $R_{\rm max}/h_{\rm 
R} = 3.6 \pm 0.6$ (1$\sigma$) in the {\it I} band. Previous
studies have also revealed the relation between the truncation radius
and disc scalelength, but yielded different values for the ratio
$R_{\rm max}/h_{\rm R}$. For example, for 7 nearby spirals KS3 find a
mean value $R_{\rm max}/h_{\rm R} = 4.2 \pm 0.5$, \citet*{PDL00} find $2.9
\pm 0.7$ from 31 galaxies, whereas \citet{SD00} obtain $3.7 \pm 0.9$
from 61 galaxies (although the latter two samples have several
galaxies in common). These differences may be partly due to the use of
different passbands; the use of bluer passbands will result in larger
scalelengths, and hence leads to smaller $R_{\rm max}/h_{\rm R}$.

We further investigate the distribution of $R_{\rm max}/h_{\rm R}$ in
Fig.~\ref{fig:trunc}b by plotting $R_{\rm max}/h_{\rm R}$ as a
function of scalelength. For those galaxies for which no truncation
was found, a lower limit is shown (arrows). The plot seems to reveal
an increase in $R_{\rm max}/h_{\rm R}$ towards small scalelengths; the
average ratio for the spirals with $h_{\rm R} < 4$ kpc is 4.4, more 
than one standard deviation higher than that of the entire sample. 
This increase of $R_{\rm max}/h_{\rm R}$ may be related to its decrease
found at very large scalelengths by \citet{PDL00}. However, the
reality of this feature is not clear considering the modest sample
size and the selection effect against galaxies of small physical
sizes. In addition, the ability to find a truncation depends, among 
other things, on the edge-on central surface brightness
(Sect.~\ref{sec:trunc_results}). We are, for example, still missing low
surface brightness galaxies, which may have entirely different $R_{\rm
max}/h_{\rm R}$.

While our view of the distribution of $R_{\rm max}/h_{\rm R}$ is
certainly obscured by these complex selection effects, additional
information can be obtained from face-on samples. We have estimated a
{\it minimum} $R_{\rm max}$ for the spirals in the face-on sample of
\citet{JK94}. By taking the lowest contour in their {\it R}-band
contour plots, we find an average $\left(R_{\rm max}/h_{\rm
R}\right)_{\rm min} = 4.0 \pm 1.1$ (1$\sigma$). When combined with the
results from our truncation analysis, this suggests that the ratio of
truncation radius to disc scalelength in small scalelength spirals
is at least four. This is important, because small spirals are the
most numerous in the local Universe (\citealt{K87}; \citealt{JIII96},
Sect.~\ref{sec:flat}). Note that this also implies that the truncation
is more difficult to detect in small scalelength spirals.

We can compare the results on the truncation radii to the predictions
of the collapse model of disc galaxy formation
\citep{WR78,FE80,DSS97}. The collapse theory is compelling because its
predictions are in general agreement with the basic structure and
rotation curves of disc galaxies, over the full range of central
surface brightness (\citealt{DSS97}; \citealt*{MMW98}). If it is
further assumed that proto galaxies are sharp-edged then the collapse
theory also predicts the outermost radius of the baryonic disc, which
corresponds to the material with the highest specific angular
momentum. To quantify that prediction and make a comparison with the
observed truncation radii of stellar discs, we calculated model
surface density profiles over a range in total mass, $M_{\rm tot}$ =
10$^{10}$--10$^{13}$ M$_{\sun}$, and spin parameter, ${\rm \lambda}$ =
0.01--0.28, using the method of \citet{DSS97}. In the calculations, we
used a fixed baryonic mass fraction ${\rm F=0.10}$, and a Hernquist
halo at the onset of baryon-decoupling \citep[as in][]{DSS97}. From
the calculated surface density profiles, the scalelengths were derived
using a method similar to the `marking-the-disc' method \citep{F70}
and the outermost radii were obtained by taking the radius at which
the density drops to zero. The result is plotted in
Fig.~\ref{fig:trunc}b as lines of constant mass. Both the outermost
radius and the scalelength of the baryonic proto-disc increase with
the mass and angular momentum of the proto galaxy, such that their
ratio remains roughly constant at 3--4. Taking a different or
non-constant baryon mass fraction does not significantly change this
result. The predicted ratio is somewhat smaller than the ratio of 4.5,
predicted by \citet{K87} based on a comparison of the angular momentum
distribution of an exponential disc to that of a uniformly rotating,
uniform sphere with $\lambda = 0.07$.

The predictions of the collapse model of disc galaxy formation roughly
coincide with the observed truncation radii of stellar discs, although
they do not show an increase of $R_{\rm max}/h_{\rm R}$ toward small
scalelengths. It is clear from Fig.~\ref{fig:trunc}b that some
galaxies fall above or below the predicted range, which may be
attributed perhaps to environmental effects such as accretion and
mergers. At any rate, in general the ratio $R_{\rm max}/h_{\rm R}$
predicted by the collapse model is similar to that seen in the stellar
luminosity density. This may indicate that the present radial
structure of the stellar disc still largely traces that of the
baryonic proto-disc. If true, this result implies that H\,{\sc i}
present beyond the stellar disc \citep{BR97} is of a more recent 
origin \citep{K87}. The result does not seem to be in accordance with
the radial colour gradients observed in nearby face-on spirals. These
indicate a stellar population change, with both the average stellar
age and metallicity decreasing towards larger radii
\citep{JIV96}. However, whether this shows that spiral galaxies have
grown since their formation is largely an open question, for it is
unclear if the radial colour gradient applies to the entire stellar
disc or is caused by a young population in the plane of the galaxy.

It should be noted that our observations compare equally well to the
value predicted by the theory of stochastic self-propagating star
formation, for which, assuming a flat rotation curve, $R_{\rm
max}/h_{\rm R}$ = 4 \citep{SSE84}. The other possible scenario, in
which the truncation of the stellar disc is caused by a threshold on
star formation due to its dynamical stability \citep{FE80,Ken89}, does
not make a direct prediction for $R_{\rm max}$. However, in a disc
galaxy formation model similar to that of \citet{DSS97}, \citet{vdB01}
has incorporated a star formation recipe which includes a threshold
surface density based on Toomre's (\citeyear{T64}) criterion for local
stability. For models of small bulge-to-disc ratio and with maximum
rotational velocities exceeding 100 km s$^{-1}$, comparable to the
galaxies considered here, the presence of this threshold produces
truncations in the stellar discs with $R_{\rm max}/h_{\rm R}$ = 2--4
(his fig.~7). This is in good agreement with the observations and also
corresponds to the outermost radius of the proto-disc according to the
collapse model investigated above. In particular, this latter
coincidence seems to be caused by the fact that for massive late-type
spirals the model of \citet{vdB01} predicts that star formation occurs
essentially throughout the entire gas disc, such that the present day
truncation radius of the stellar disc is only marginally smaller than
the outermost radius of the gaseous disc. We tentatively conclude that
although the origin of the truncation of the stellar discs is still
unclear, and may well be caused by a threshold on star formation, the
truncation radius mainly reflects the material of the highest specific
angular momentum which has collapsed during formation, at least for
the intermediate to late-type galaxies considered here. To test
whether the stellar truncation radius corresponds to the predicted
threshold radius on star formation would require a detailed study of
H\,{\sc i} and H$_{2}$ surface density profiles, as first suggested by
\citet{Ken89}.

\section{Conclusions}
\label{sec:conclusions}

We have analysed {\it I}-band photometry of 34 edge-on spiral
galaxies, taken from the sample of \citet{G98}, to study the
flattening and truncation of stellar discs. The exponential scale
parameters of the disc light were derived using a 2D least-squares
fitting method. This procedure includes a bulge-disc decomposition and
adopts a mask in the region $|z| \leq 1.5\ h_{\rm z}$ to systematically
exclude the region affected by dust extinction and young
populations. It is better suited to extract the disc scalelength than
the earlier 1D analysis \citep{G98}. We used an edge-fitting method,
similar to the one introduced by \citet{GKW01}, to study the disc
truncation. This method uses a simple straight line as a fitting
function and a quantitative criterion to establish the presence of a
truncation.

We find a clear increase in the scaleheight of the stellar disc as a
function of maximum rotational velocity and total H\,{\sc i}
mass. This is in general accordance with observations of the stellar
kinematics in spiral galaxies \citep{B93}; larger discs are clearly more
dynamically evolved and thicker.
Both, the maximum and the range of the flattening of stellar discs
seem to increase with maximum rotation and total H\,{\sc i} mass. We
used the average volume corrected flattening of the disc light,
$\left<h_{\rm R}/h_{\rm z}\right> = 7.3$, to estimate the disc
contribution to the rotation curve, with the assumption that the
flattening of the disc mass equals that of the disc light. This
resulted in a disc contribution of 57$\pm$22 percent, which still
-- barely -- allows for a maximum disc situation.

At least 20 galaxies in our sample (60 percent) are truncated,
displaying a tight relation between scalelength and truncation
radius. The average truncation radius corresponds to 
$3.6\pm0.6$ {\it I}-band scalelengths. In addition, there seems to be
a small increase in $R_{\rm max}/h_{\rm R}$ towards galaxies of small
scalelength. Small scalelength spirals, which are the most numerous
spirals in the local Universe, have an $R_{\rm max}/h_{\rm R}$ of at
least four. A comparison of this ratio to that predicted by theories
of disc galaxy formation seems to indicate that for the intermediate
and late-type galaxies considered here the truncation radius mainly
reflects the material of the highest specific angular momentum which
has collapsed during formation. These scenarios are inconclusive with
respect to the question of the origin of the truncation, and imply
that a structure analysis of the truncation alone is insufficient to
solve this issue.

It is clear that many questions are left unanswered. In an attempt to
partly remedy this situation, we are currently studying the stellar
and H\,{\sc i} kinematics of the galaxies in our sample. This will,
for instance, further improve the statistics on the empirical relation
between maximum rotation and the stellar velocity dispersion, which
could significantly advance our knowledge on the dark matter content
of spiral galaxies. Still, studies of larger samples are needed to
further constrain many of the relations found here, e.g. the trend of
the disc flattening with disc mass. These future studies should
preferably use deep, wide field, near infrared photometry. By
targeting face-on as well as edge-on spirals, it becomes possible
to more accurately study the disc truncation as well as its shape.

\section*{Acknowledgments}
We would like to thank Reynier Peletier and the anonymous referee for
useful suggestions. MK would like to express his sincere thanks to
Roelof Bottema for his sharp comments on an earlier version of this
paper. RdeG acknowledges partial support from NASA grants NAG 5-3428
and NAG 5-6403, and hospitality at the Kapteyn Astronomical Institute
on 2 visits. This work is based on observations obtained at the
European Southern Observatory, La Silla, Chile. We have made use of
the LEDA database (http://leda.univ-lyon1.fr).

\bibliographystyle{mn}
\bibliography{mn-jour,article}

\section*{Appendix}

We further illustrate the results of the 2D fits
(Sect.~\ref{sec:sb_results}) and the truncation analysis
(Sect.~\ref{sec:trunc_results}). For each galaxy, information is
presented in three panels.

The panel at the top left shows the surface brightness distribution in
the {\it I} band, after editing and rotation
(Sect.~\ref{sec:sb_results}). The surface brightness at the lowest
contour corresponds to the 3$\sigma$ noise level of the 
background. The 3$\sigma$ level is marked by the solid arrow on
the magnitude scale of the lower left panel. The contour interval is
0.5 mag arcsec$^{-2}$, except for ESO 033-G22, ESO 041-G09 and ESO
138-G14, where we chose an interval of 1.0 mag arcsec$^{-2}$ for
clarity. The orientation of each image is given in
Table~\ref{tab:appendix}, columns (5) \& (6). The inner and outer
radial fitting boundaries and the mask (Sect.~\ref{sec:sb_results})
are marked by the vertical and horizontal dotted lines
respectively. The dashed contour marks the boundary outside of which
the data were not included in the 2D fit (Sect.~\ref{sec:sb_approach}).

The bottom left panel shows, on the same scale as the top left image,
two profiles taken parallel to the major axis. In each case the
best-fitting model is shown for comparison (dashed lines). The bottom
profile (solid line) is the average of two profiles. These were taken
at a distance of 1.5 ${h_{\rm z}}$ on each side of the galactic plane
and averaged over a range of 1/3 ${h_{\rm z}}$ to yield an acceptable
signal-to-noise. When one side of the galaxy was fitted, only that
side is used to construct this profile. The inner and outer radial
fitting boundaries are indicated by the horizontal dotted lines. The
top profile in this panel is the one used in the truncation
analysis. It is shifted upwards by 1.5 magnitudes for clarity. If a
truncation is present then the truncation fit (solid line) is also
shown. The dashed--dotted lines indicate an envelope corresponding to
$\pm 3$ times the background error. The open arrow at the magnitude
scale, also shifted by 1.5 mag, indicates the standard deviation of
the background in the vertically averaged profile used to clip the
light profile (see Sect.~\ref{sec:trunc_results} for further
details).

The panel at the bottom right shows two profiles taken perpendicular
to the major axis. The corresponding projected radii at which these
were taken are mentioned in Table~\ref{tab:appendix}, columns (7) \&
(8). In each case, the model is shown for comparison (dashed
lines). Again, each of the profiles is an average of two profiles, one
taken on each side of the galaxy after averaging over a range of 1/3
$h_{\rm R}$. The boundary of the mask is indicated by the dotted
lines.

\tabcolsep=0.95mm
\begin{table}
\caption[]{{\bf Additional properties of the 2D fits presented in the
appendix}\\
Columns: (1) Name (ESO-LV), the galaxies for which a mask $|z| \leq 1.0\
h_{\rm z}$ was used are indicated by an asterisk; (2) and (3) Centre
used in the 2D fits; (4) Adopted major axis position angle (N
$\rightarrow$ E) (5) and (6) Image orientation; (7) and (8) Radii used
to extract the vertical profiles.}
\begin{tabular}{lccrllcc}
\hline
Galaxy & R.A. & Dec. & P.A. & Left & Top & $R_{1}$ & $R_{2}$\\
       & ($^{\rm h}$ $^{\rm m}$ $^{\rm s}$) & ($^{\rm d}$ $^{\rm m}$
$^{\rm s}$) & (deg) & &     & ($h_{\rm R}$) & ($h_{\rm R}$)\\
(1)    & (2) & (3) & (4) & (5) & (6) & (7) & (8)\\
\hline
026-G06 & 20 48 27.9 & -78 04 09.1 & 76.1 & W & S & 0.5 & 1.5\\
033-G22 & 05 31 41.7 & -73 45 06.1 & 168.1 & N & W & 1.2 & 2.5\\
041-G09 & 14 47 43.8 & -73 18 21.5 & 132.7 & SE & NE & 0.7 & 1.5\\
138-G14 & 17 06 59.4 & -62 04 59.4 & 135.1 & SE & NE & 0.5 & 1.5\\
141-G27 & 19 07 06.7 & -59 27 59.0 & 125.1 & SE & NE & 0.5 & 1.4\\
142-G24 & 19 35 42.2 & -57 31 05.2 & 7.2 & S & E & 0.7 & 2.4\\
157-G18 & 04 17 54.6 & -55 55 54.4 & 18.9 & NE & NW & 0.5 & 1.5\\
201-G22$^{*}$ & 04 08 59.9 & -48 43 37.3 & 59.1 & NE & NW & 0.7 & 1.5\\
202-G35 & 04 32 16.5 & -49 40 35.0 & 134.2 & NW & SW & 0.5 & 1.5\\
240-G11 & 23 37 49.5 & -47 43 38.1 & 127.6 & SE & NE & 0.8 & 1.9\\
263-G15 & 10 12 19.8 & -47 17 40.4 & 108.6 & NW & SW & 0.5 & 1.8\\
263-G18$^{*}$ & 10 13 30.5 & -43 43 00.0 & 128.5 & NW & SW & 0.2 & 1.5\\
269-G15$^{*}$ & 12 57 13.1 & -46 22 40.8 & 89.0 & N & W & 0.7 & 2.0\\
288-G25 & 21 59 17.8 & -43 52 01.3 & 54.1 & SW & SE & 0.5 & 2.0\\
315-G20$^{*}$ & 09 42 22.0 & -41 48 56.7 & 58.9 & NE & NW & 0.5 & 1.5\\
321-G10 & 12 11 42.1 & -38 32 54.4 & 71.3 & NE & NW & 1.0 & 1.7\\
322-G87$^{*}$ & 12 48 02.7 & -40 49 06.5 & 138.6 & NW & SW & 0.5 & 1.5\\
340-G08 & 20 17 11.6 & -40 55 28.4 & 34.9 & SW & SE & 1.5 & 2.5\\
340-G09$^{*}$ & 20 17 21.1 & -38 40 26.7 & 98.5 & E & N & 0.7 & 1.5\\
416-G25 & 02 48 40.8 & -31 32 09.5 & 25.1 & NE & NW & 1.1 & 2.0\\
435-G14 & 09 57 48.2 & -28 30 23.6 & 54.5 & NE & NW & 1.0 & 2.0\\
435-G25 & 09 59 55.4 & -29 37 02.4 & 77.4 & NE & NW & 0.1 & 1.2\\
435-G50 & 10 10 50.2 & -30 25 25.7 & 70.2 & NE & NW & 0.5 & 2.0\\
437-G62$^{*}$ & 10 48 04.4 & -31 32 00.5 & 179.5 & N & W & 0.5 & 1.5\\
446-G18 & 14 08 38.3 & -29 34 19.4 & 7.2 & S & E & 0.5 & 1.5\\
446-G44 & 14 17 49.1 & -31 20 56.3 & 77.1 & W & S & 0.0 & 1.0\\
460-G31$^{*}$ & 19 44 21.5 & -27 24 24.6 & 92.4 & E & N & 0.5 & 1.5\\
487-G02 & 05 21 48.2 & -23 48 35.5 & 60.6 & NE & NW & 0.5 & 2.5\\
506-G02$^{*}$ & 12 20 10.1 & -26 04 00.4 & 4.2 & N & W & 0.2 & 1.2\\
509-G19 & 13 27 56.3 & -25 51 22.4 & 51.3 & SW & SE & 0.7 & 1.2\\
531-G22 & 21 40 29.5 & -26 31 39.3 & 9.3 & S & E & 0.7 & 1.5\\
555-G36 & 06 07 41.9 & -19 54 45.2 & 145.5 & NW & SW & 1.0 & 2.0\\
564-G27 & 09 11 54.7 & -20 07 02.4 & 167.6 & N & W & 0.5 & 1.8\\
575-G61$^{*}$ & 13 08 15.4 & -21 00 06.1 & 175.3 & N & W & 0.0 & 2.0\\
\hline
\end{tabular}
\label{tab:appendix}
\end{table}

\begin{figure*}
\centering
\resizebox{7.5cm}{!}{\includegraphics{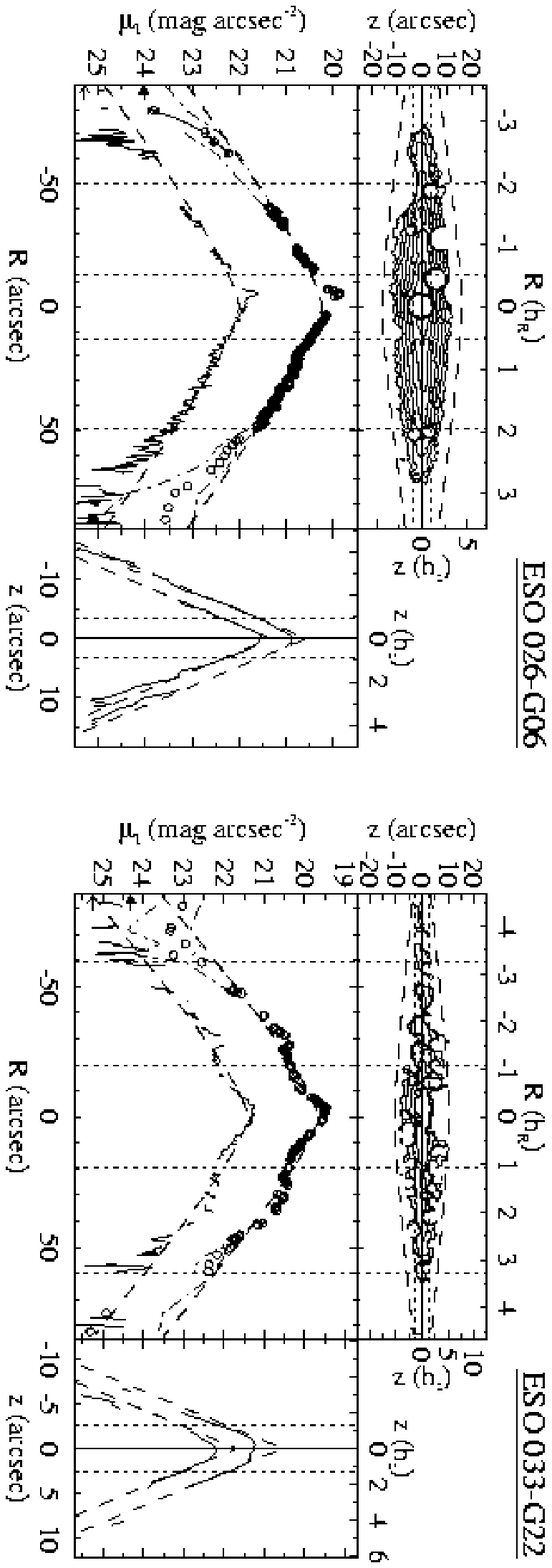}}
\end{figure*}
\begin{figure*}
\centering
\resizebox{15.0cm}{!}{\includegraphics{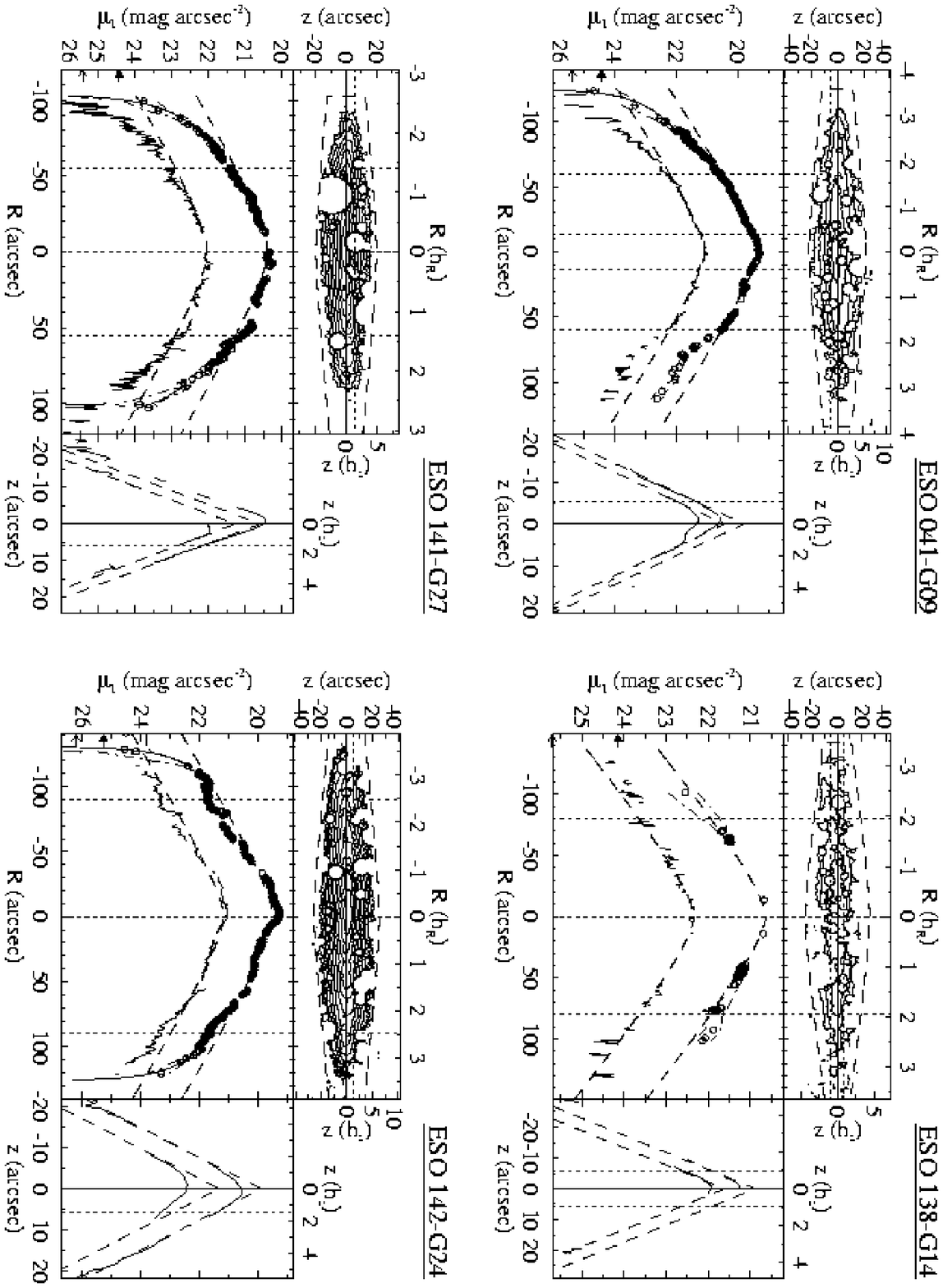}}
\end{figure*}
\begin{figure*}
\centering
\resizebox{15.0cm}{!}{\includegraphics{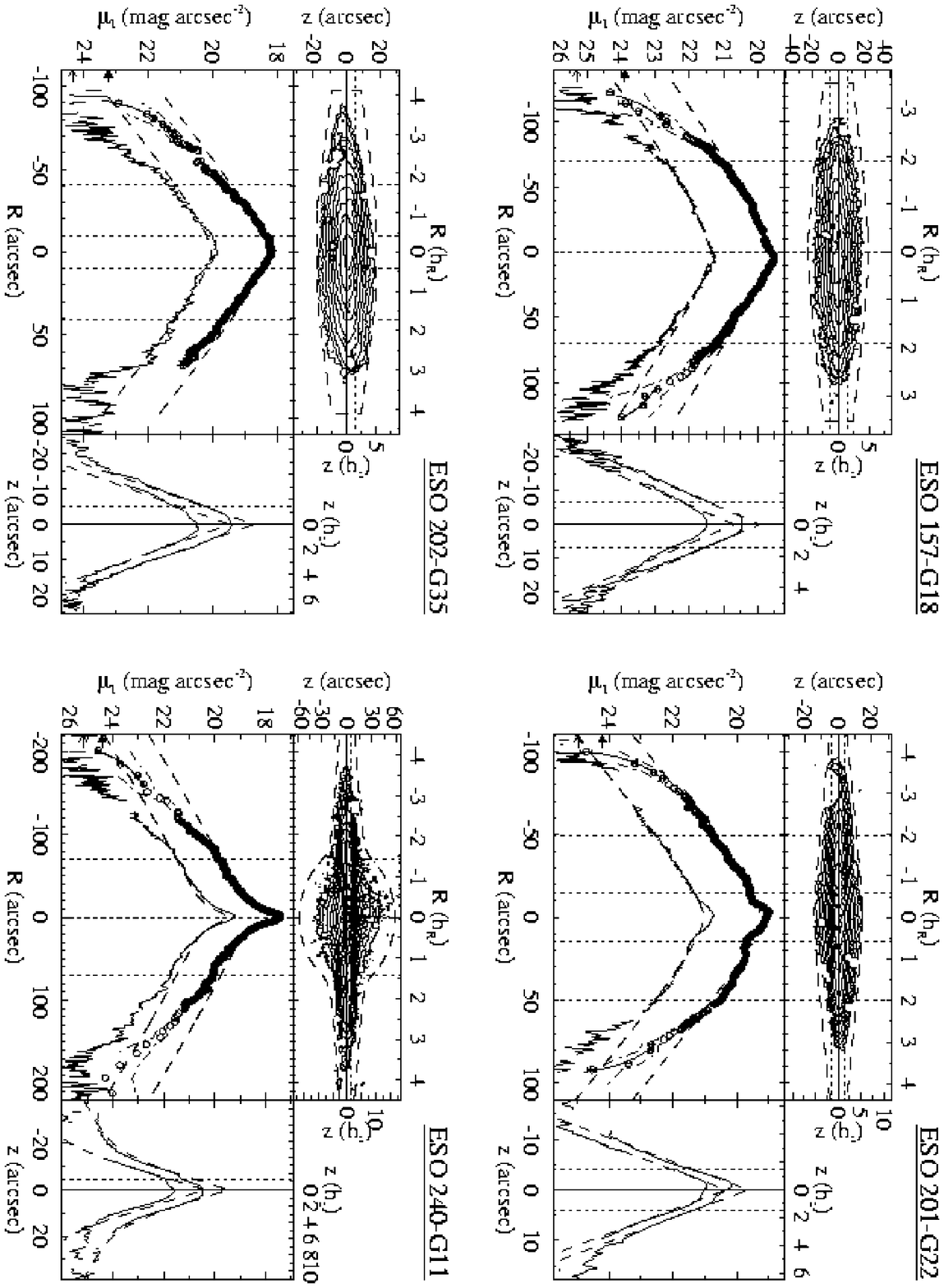}}
\end{figure*}
\begin{figure*}
\centering
\resizebox{15.0cm}{!}{\includegraphics{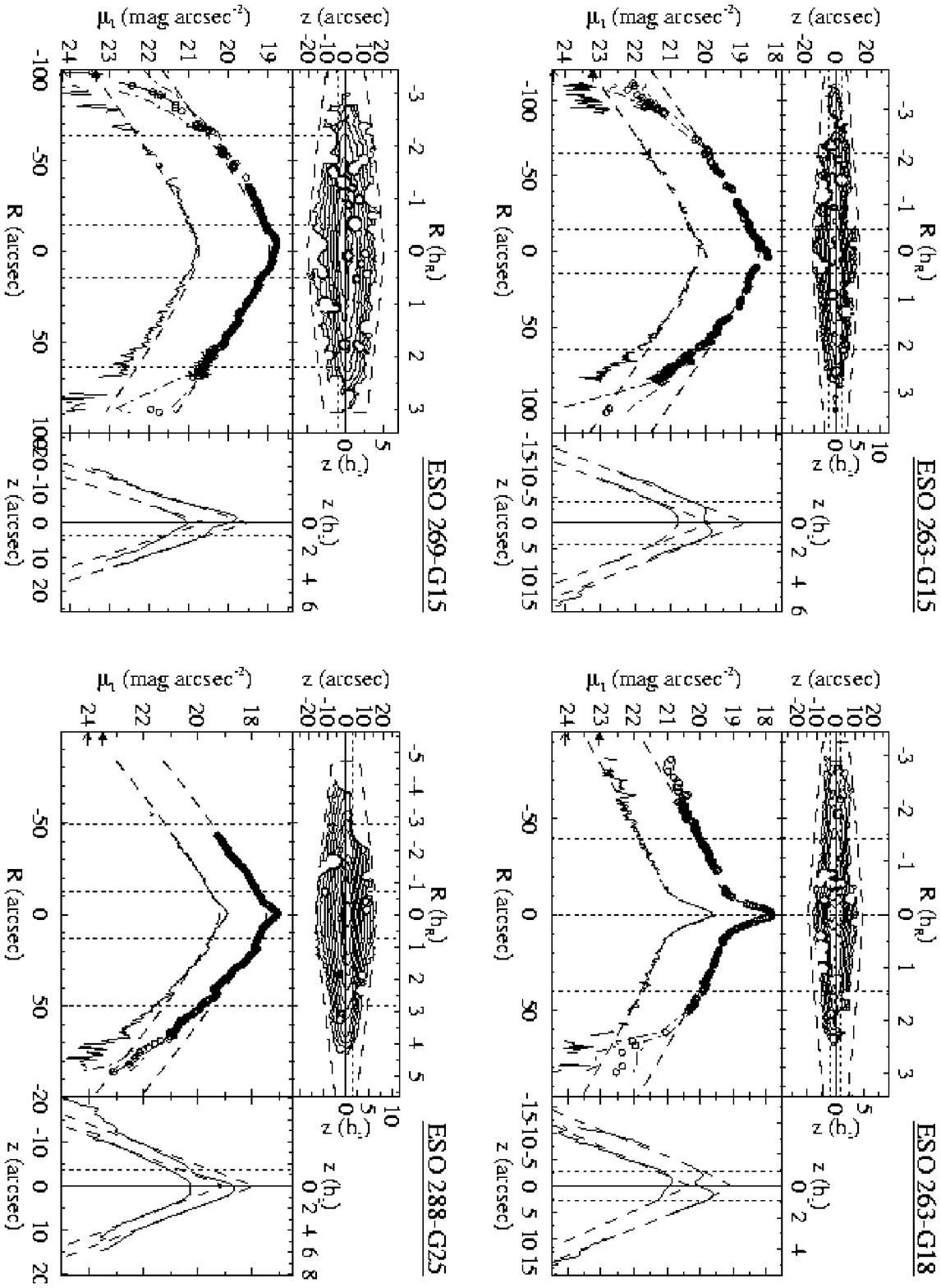}}
\end{figure*}
\begin{figure*}
\centering
\resizebox{15.0cm}{!}{\includegraphics{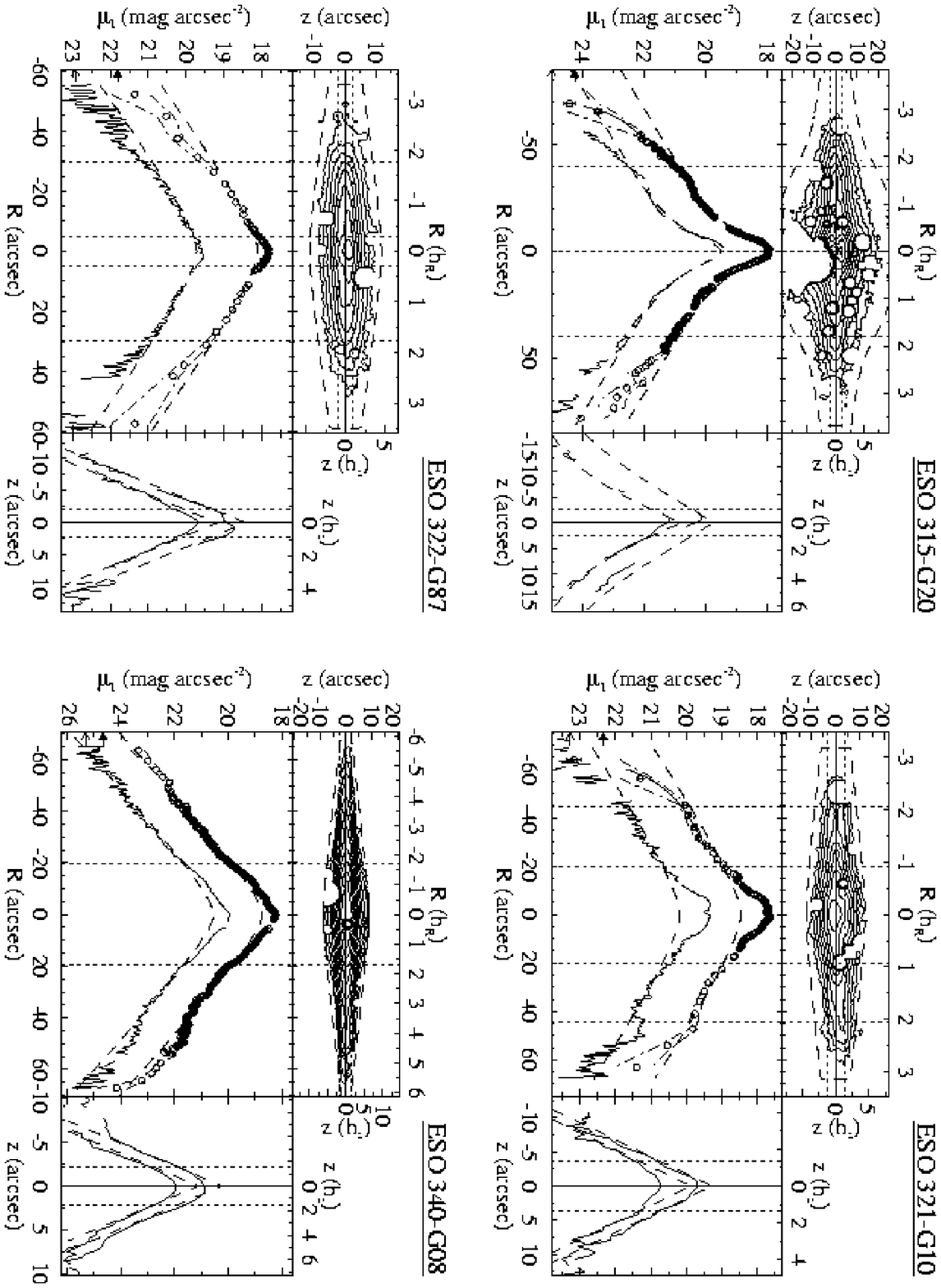}}
\end{figure*}
\begin{figure*}
\centering
\resizebox{15.0cm}{!}{\includegraphics{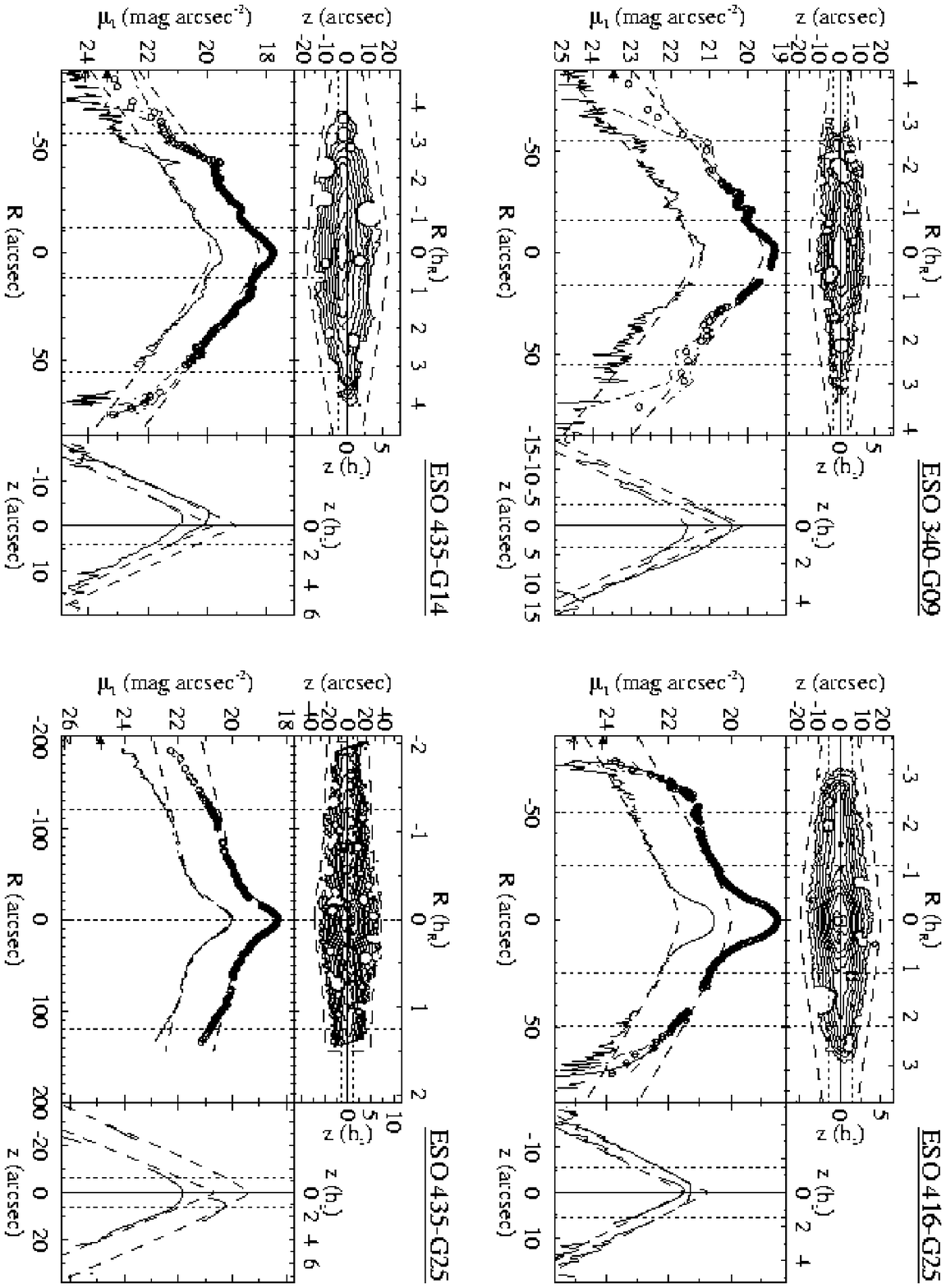}}
\end{figure*}
\begin{figure*}
\centering
\resizebox{15.0cm}{!}{\includegraphics{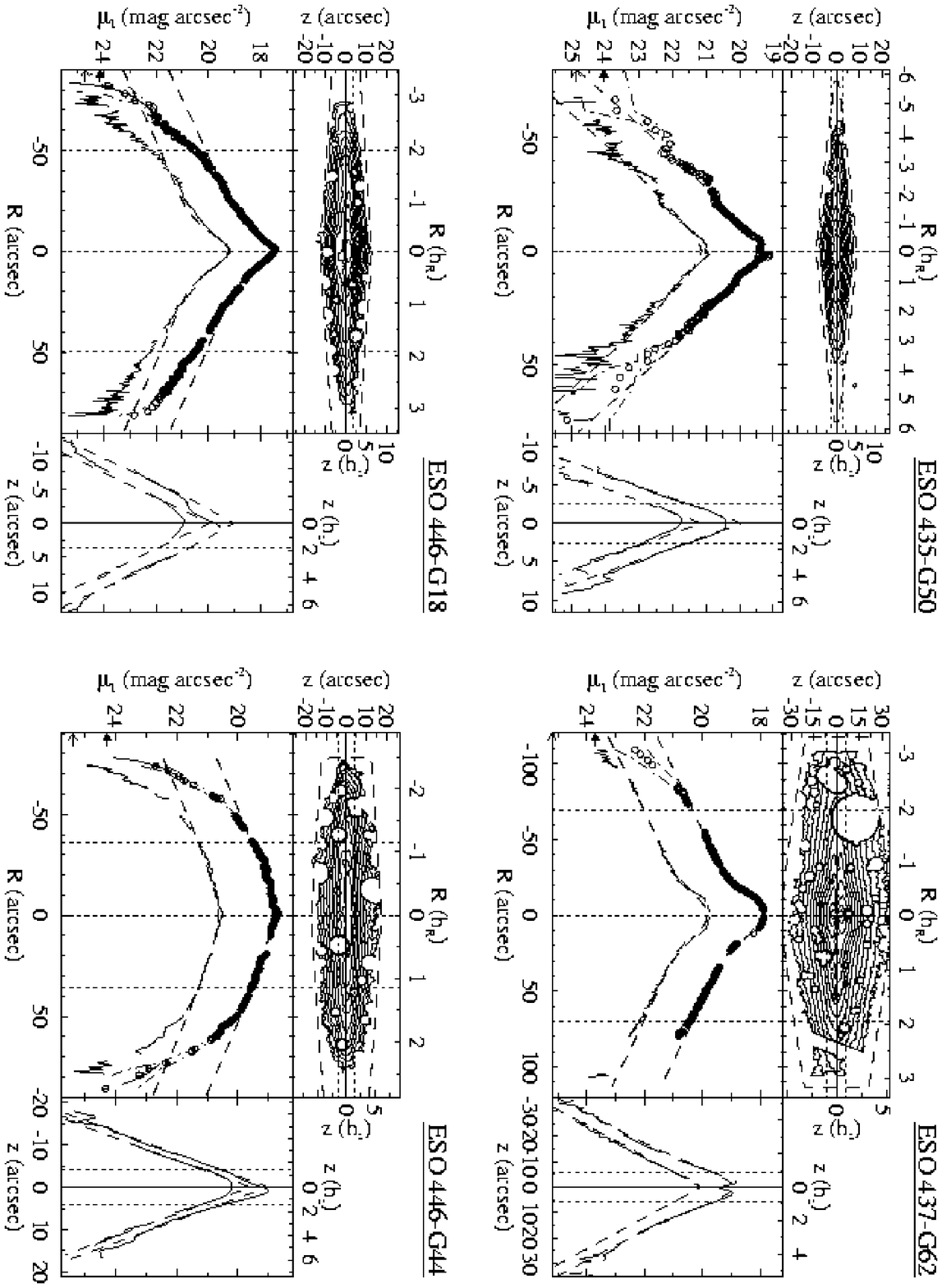}}
\end{figure*}
\begin{figure*}
\centering
\resizebox{15.0cm}{!}{\includegraphics{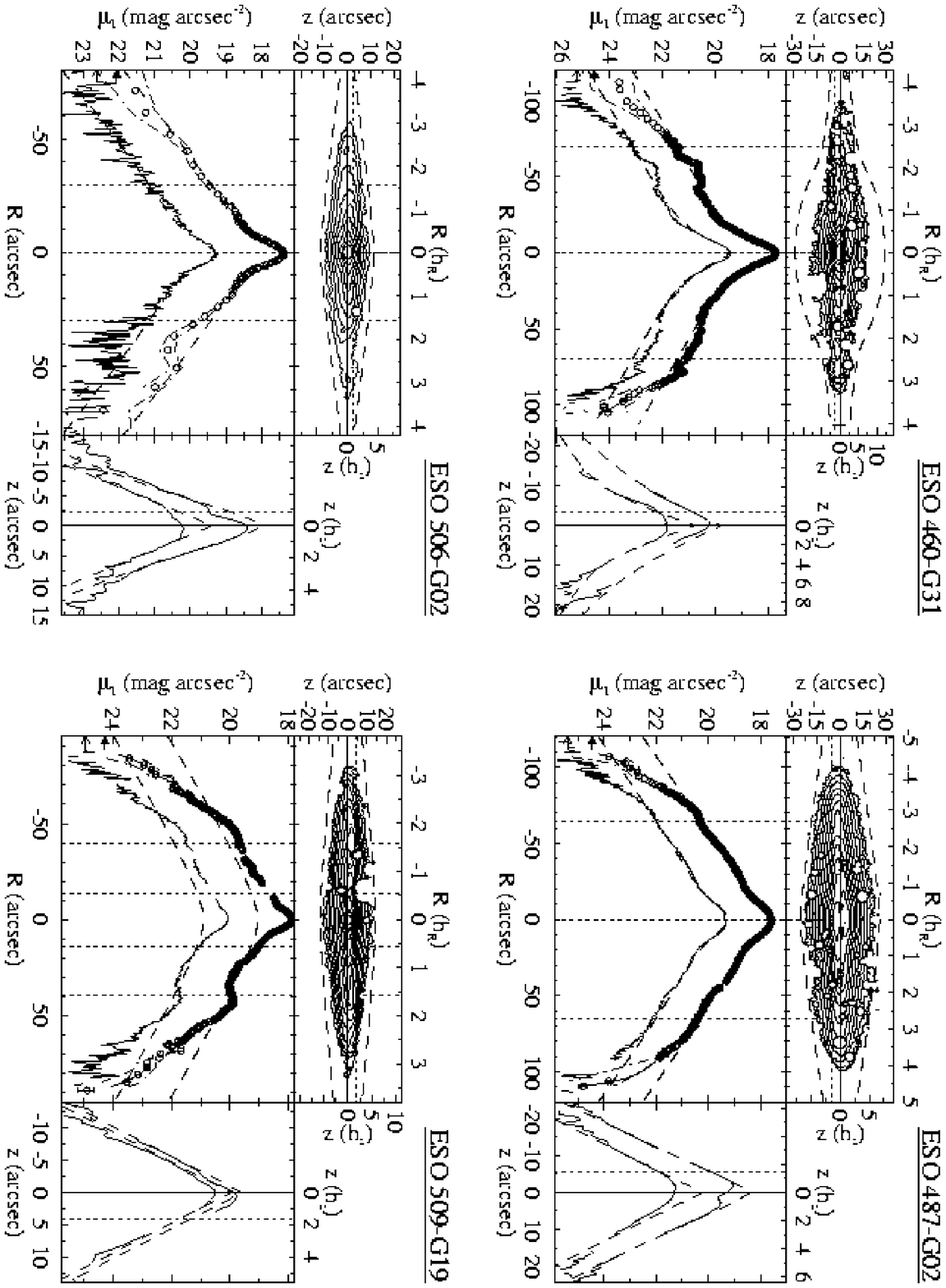}}
\end{figure*}
\begin{figure*}
\centering
\resizebox{15.0cm}{!}{\includegraphics{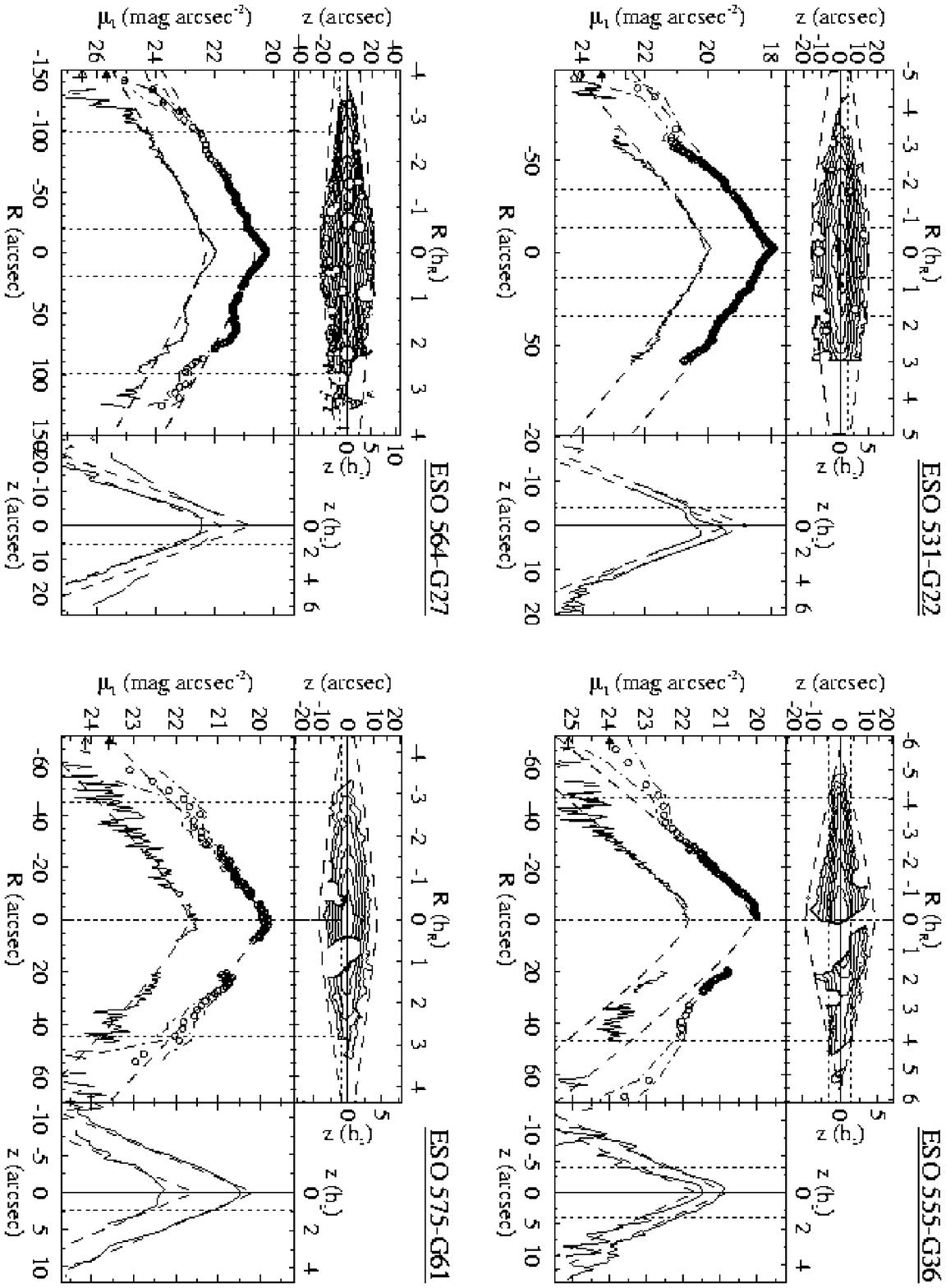}}
\end{figure*}

\end{document}